\documentclass{elsarticle}

\usepackage{lineno}
\modulolinenumbers[5]

\usepackage{graphicx}        

\usepackage{lscape}

\usepackage{amsmath}

\newcommand{\be}{\begin{equation}}
\newcommand{\ee}{\end{equation}}
\newcommand{\bra}{\langle}
\newcommand{\ket}{\rangle}
\newcommand{\bea}{\begin{eqnarray}}
\newcommand{\eea}{\end{eqnarray}}

\begin{document}

\begin{frontmatter}
\title{Statistical properties and multifractality of Bitcoin}

\author{Tetsuya Takaishi\corref{mycorrespondingauthor}}
\address{Hiroshima University of Economics, Hiroshima 731-0192 JAPAN}
\cortext[mycorrespondingauthor]{Corresponding author}

\ead{tt-taka@hue.ac.jp}

\begin{abstract}
Using 1-min returns of Bitcoin prices,
we investigate statistical properties and multifractality of a Bitcoin time series.
We find that the 1-min return distribution is fat-tailed, and kurtosis largely deviates from the Gaussian expectation.
Although for large sampling periods, kurtosis is anticipated to approach the Gaussian expectation,
we find that convergence to that is very slow.
Skewness is found to be negative at time scales shorter than one day and becomes consistent
with zero at time scales longer than about one week.
We also investigate daily volatility-asymmetry by using GARCH, GJR, and RGARCH models, and
find no evidence of it.
On exploring multifractality using multifractal detrended fluctuation analysis,
we find that the Bitcoin time series exhibits multifractality.
The sources of multifractality are investigated, confirming that both temporal correlation and the fat-tailed distribution
contribute to it.
The influence of "Brexit" on June 23, 2016 to GBP--USD exchange rate and Bitcoin is examined in multifractal properties. We
find that, while Brexit influenced the GBP--USD exchange rate, Bitcoin was robust to Brexit.  
\end{abstract}


\end{frontmatter}


\section{Introduction}
In 2008, Bitcoin was first proposed as a cryptocurrency in a paper by Satoshi Nakamoto\cite{Nakamoto2008}, who is still unidentified.
His proposal was quickly accepted, and at the beginning of 2009, the Bitcoin software, "Bitcoin Core," was
distributed, and a "genesis block," the first block of a block chain, was created.
Since then, the Bitcoin system has been maintained without any critical problems, proving its robustness.
The Bitcoin network is a peer-to-peer network that allows online payments without any central institutions.
The block chain used in Bitcoin is a key technology for maintaining a decentralized system;
it can be utilized not only for cryptocurrencies, but also in other financial fields.
In recent years, the importance of block chain technology has been recognized widely as a core technology of FinTech,
and extensive research has been conducted to achieve new distributed platforms for financial institutions\cite{R3,Hyperledger}.

Bitcoin price properties have also received attention, leading to analysis of various related issues, such as
time series analysis using economic models, such as GARCH-type models\cite{dyhrberg2016bitcoin,katsiampa2017volatility}; statistical properties of Bitcoin prices\cite{bariviera2017some,chu2015statistical};
inefficiency of Bitcoin\cite{urquhart2016inefficiency,nadarajah2017inefficiency}; hedging capabilities of Bitcoin\cite{dyhrberg2016hedging,bouri2017hedge}; speculative bubbles in Bitcoin\cite{cheah2015speculative};
the relationship between Bitcoin and search queries on Google trends and Wikipedia\cite{kristoufek2013bitcoin}; and so forth.
The Hurst exponent of the Bitcoin time series was also obtained using detrended fluctuation analysis (DFA)\cite{peng1994mosaic},
and was found to be time-dependent, reaching close to 0.5 after 2014\cite{bariviera2017some}.

In this paper we focus on the multifractal property of Bitcoin, which is unaddressed by current literature.
To explore the multifractal property of Bitcoin,
we calculate the generalized Hurst exponent using multifractal detrended fluctuation analysis (MF-DFA)\cite{kantelhardt2002multifractal}
by using 1-min return data.
The MF--DFA, which is an extended method of DFA, can investigate multifractal properties of non-stationary time series, and
has been successfully applied for a variety of financial markets, such as
stock \cite{matia2003multifractal,kwapien2005components,lee2006multifractal,kumar2009multifractal,zunino2009multifractal,suarez2014multifractality,hasan2015multifractal,lahmiri2017multifractal},
commodity \cite{matia2003multifractal,gu2010multifractal,li2011multifractal,mali2014multifractal,delbianco2016multifractal}, tanker \cite{zheng2016multifractal},
derivative \cite{lim2007multifractal}, foreign exchange rates\cite{norouzzadeh2006multifractal,oh2012multifractal,wang2012statistical,qin2015effectiveness,caraiani2015evidence}, 
and electricity markets\cite{norouzzadeh2007anti}.
An especially interesting application of multifractal analysis is measuring the degree of multifractality of time series, which can be
related to the degree of efficiency of financial markets\cite{zunino2008multifractal,wang2011analysis,tiwari2017multifractal,mensi2017global}.
However, whether the efficient market hypothesis (EMH)\cite{malkiel1970efficient} holds for actual markets is still
under debate. The multifractal measure could give us another insight into the EMH.
Besides multifractal analysis, we also investigate some properties of high-frequency Bitcoin time series, such as autocorrelation of (absolute) returns,
kurtosis, skewness, and volatility persistence. 
For other financial assets, these properties have been extensively investigated, and 
some stylized facts have been observed, such as "fat-tailed return distribution," 
"volatility clustering," and "long memory of absolute return"\cite{Cont2001QF}. 
We investigate whether these stylized facts hold for Bitcoin time series. 

The rest of this paper is organized as follows. Section 2 describes the data, and
presents the results of statistical properties.
In section 3, we introduce MF--DFA,
and in section 4, we present its results.
In section 5, we compare Bitcoin with GBP--USD exchange rate using multifractal analysis.
Finally, we conclude our study in section 6.

\section{Data and its Statistical Properties}
The Bitcoin data used in this study are based on a 1-min Bitcoin price index (BPI) created by Coindesk\cite{Coindesk} from January 1, 2014 to December 31,2016.
Launched in September 2013, the BPI is produced using a simple average across several Bitcoin exchanges that match Coindesk's requirements.
Although Coindesk provides the BPI before September 2013 using Mt.Gox prices as BPI,
we do not use that data in this study to avoid possible changes in the price properties, which might especially occur for high-frequency data.

Let $P(t), t=1,2,...,N$ be the time series of BPI and
the return $r(t)$ given by a logarithmic price difference, $r(t)=log P(t)-log P(t-\Delta t)$,
where $\Delta t$ is the sampling period.
Fig.1 shows the distribution of returns sampled at $\Delta t=1$-min, where the returns are normalized to a zero average with unit variance.
The return distribution clearly exhibits a fat-tailed tendency. Note that daily Bitcoin returns also show a fat-tailed tendency\cite{chu2015statistical}.

\begin{figure}
\centering
\includegraphics[height=6cm,width=8cm]{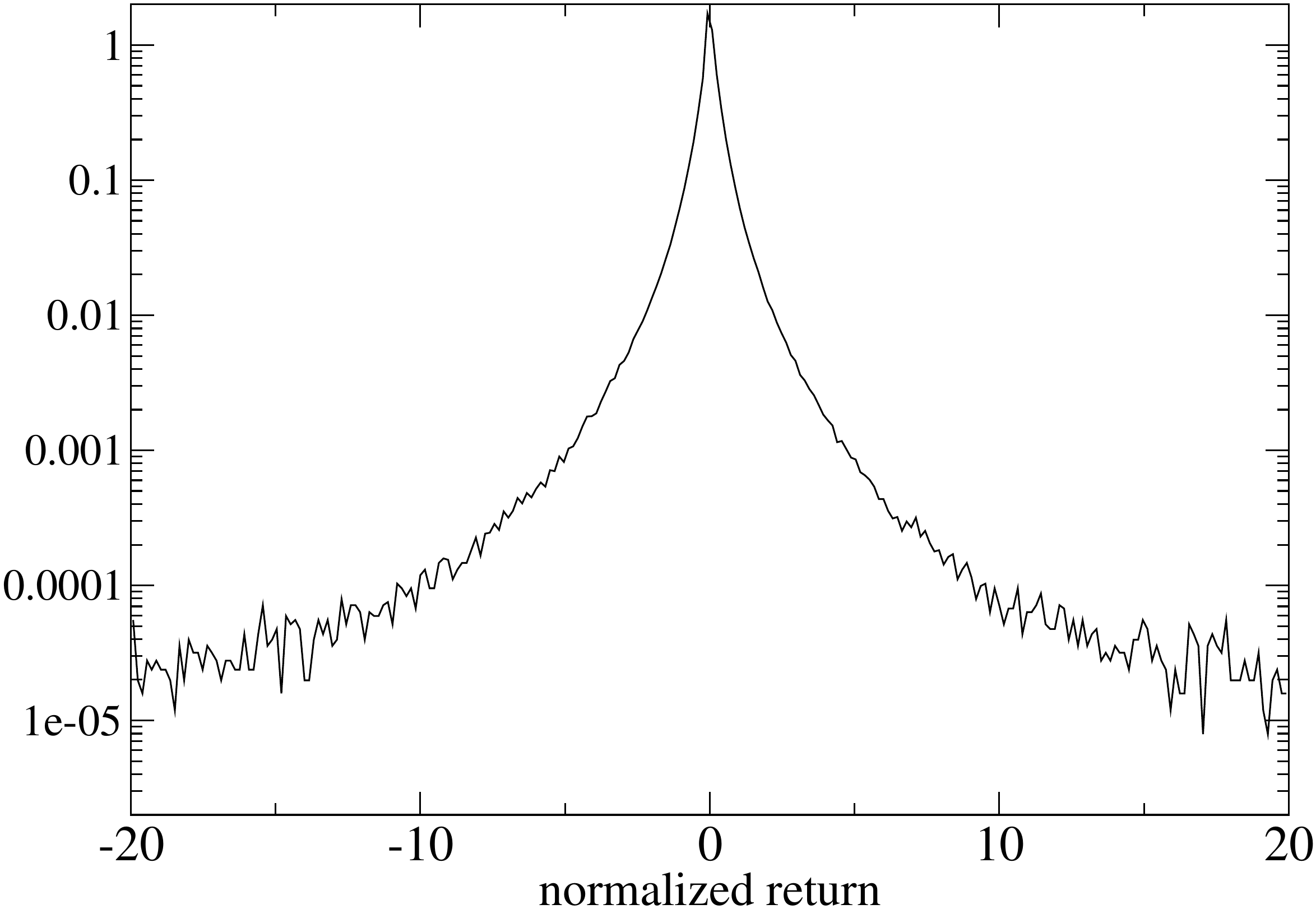}
\caption{
1-min return distribution.
}
\end{figure}

Fig.2(a)--(c) show the autocorrelation function (ACF) of returns for the periods of 2014, 2015, and 2016, respectively.
The ACF of returns quickly decreases as the time lag (min) increases, and goes to zero at around 5-min, which means that
the autocorrelation of returns is very short ranged.
Then the ACF overshoots zero, and becomes negative around 5-min to 10-min.
The similar behavior that the ACF overshoots zero is seen for the high-frequency returns of the DAX index\cite{VOIT2003286}.
At the 1-min time lag, however, we also observe a negative autocorrelation for 2014, and this might indicate that
the times series properties of 2014 differ from those of other years.
This could be related to the observation in Ref.\cite{bariviera2017some} that the Hurst exponent changes in 2014.
Fig.2(d) shows the ACF of returns for the full sample (2014--2016), and a negative autocorrelation is also observed at the 1-min time lag.

Fig.3 shows the ACF of absolute returns for the 2014--2016 period.
For the periods of 2014, 2015, and 2016, we find the results are very similar to
those of 2014--2016. Thus, we present only the results of 2014--2016 as representative.
It is seen that 
the ACF follows a power law function, indicating that Bitcoin has a long memory property in absolute returns.
After fitting the data to a power law function $\sim t^{-\mu}$, we obtain $\mu\approx 0.16$.

\begin{figure}
\centering
\includegraphics[height=4cm,width=6cm]{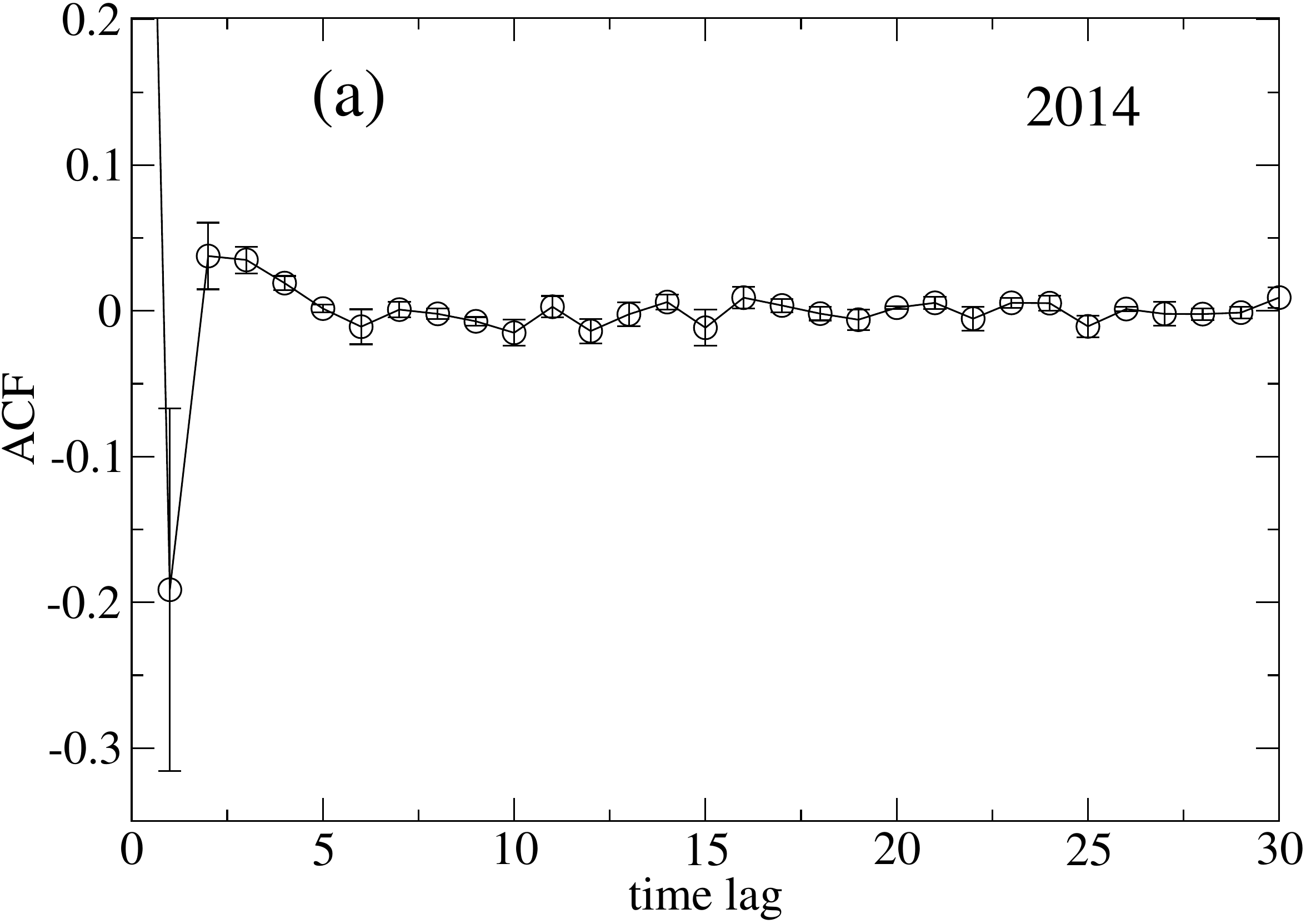}
\includegraphics[height=4cm,width=6cm]{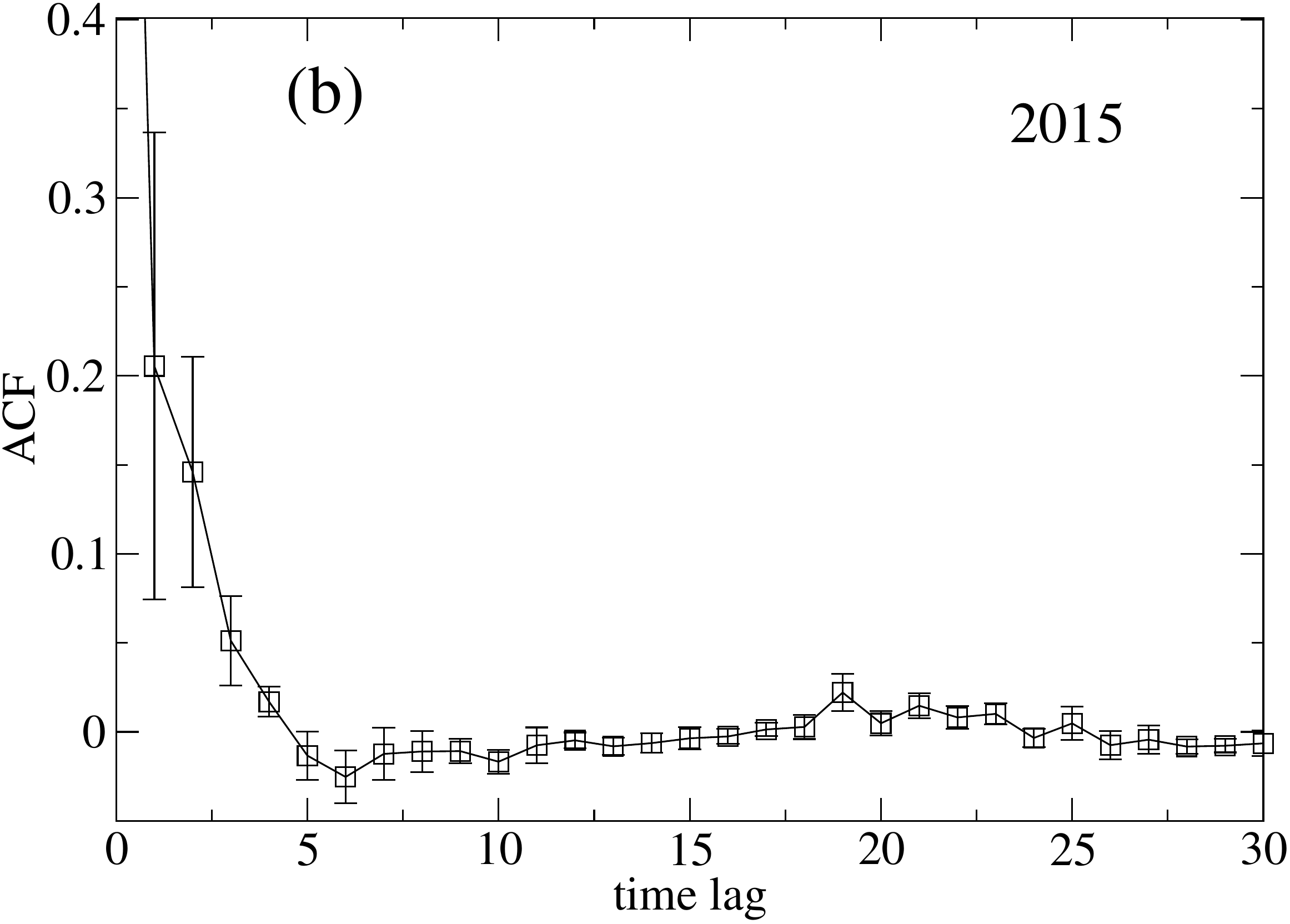}
\includegraphics[height=4cm,width=6cm]{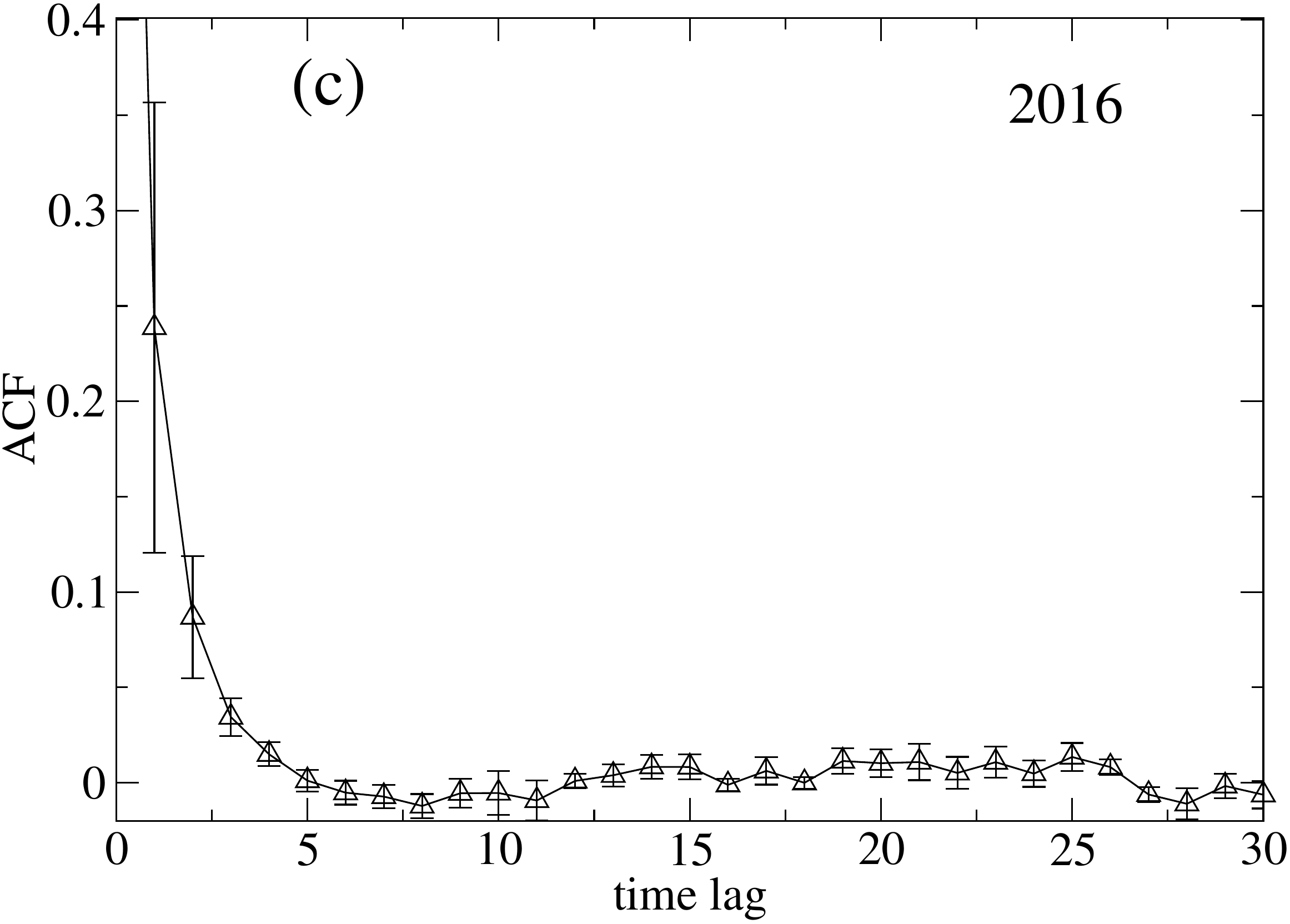}
\includegraphics[height=4cm,width=6cm]{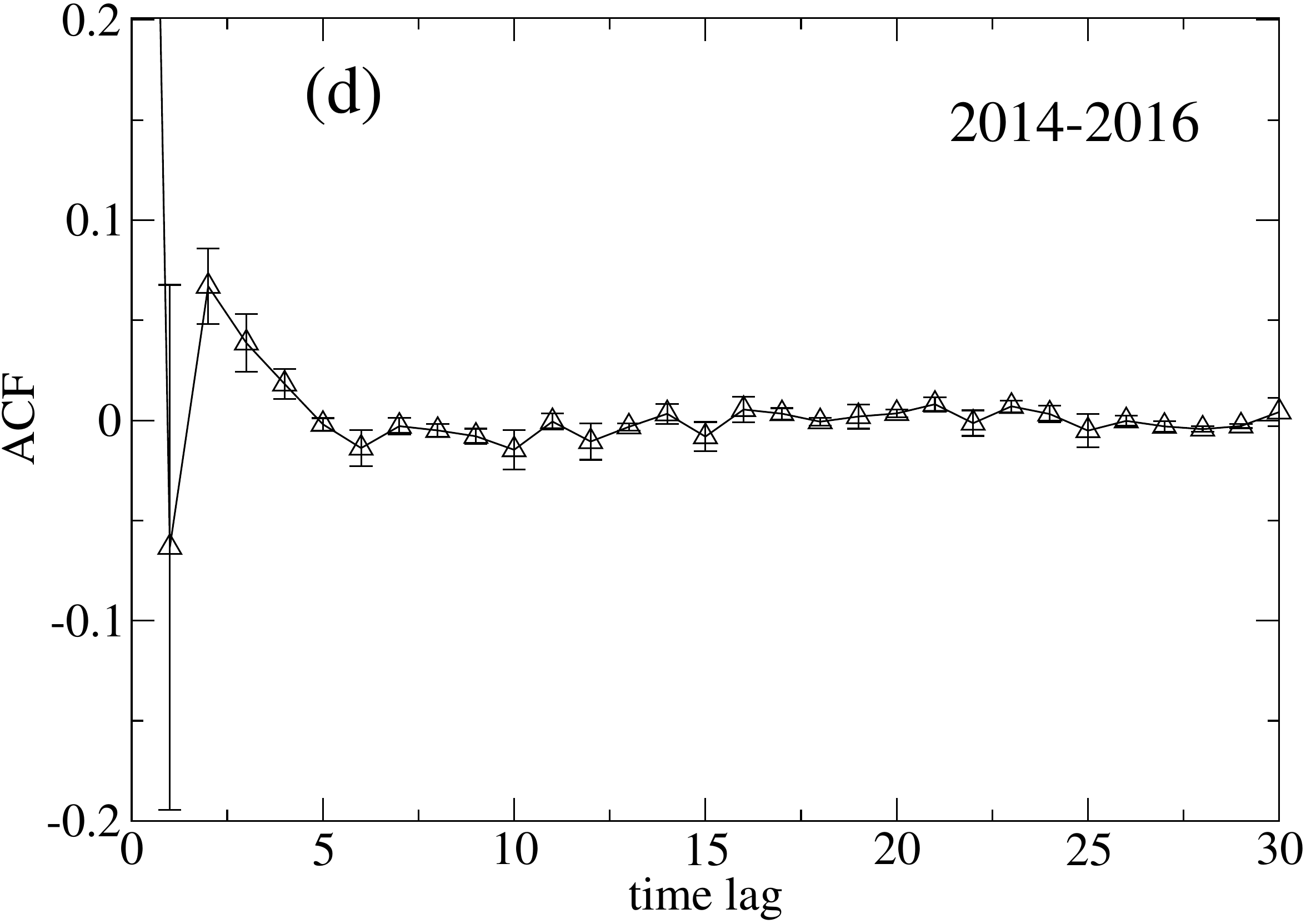}
\caption{
The ACF of returns: (a) 2014, (b) 2015, (c) 2016, and (d) 2014--2016.
}
\end{figure}

\begin{figure}
\centering
\includegraphics[height=5cm,width=8cm]{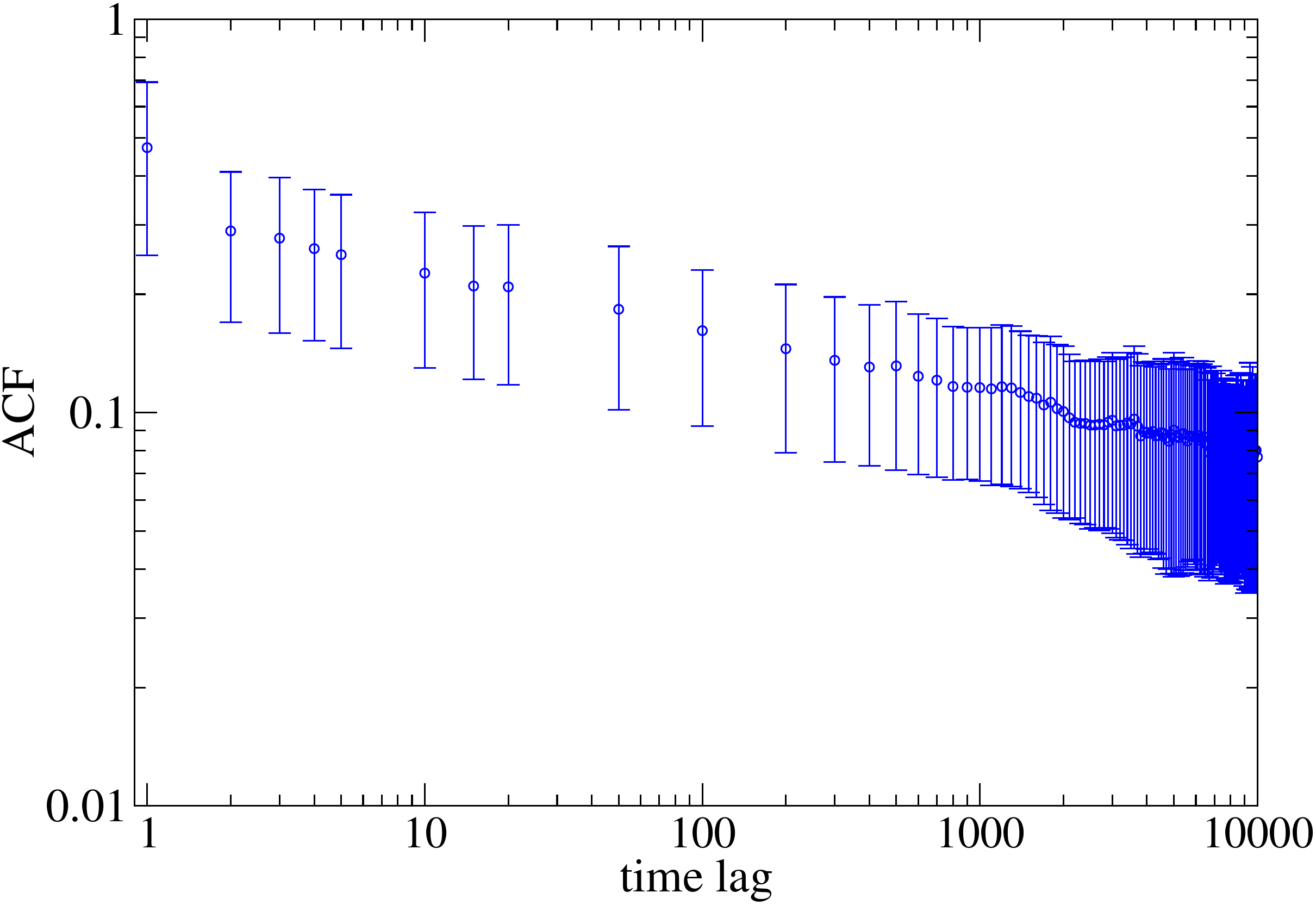}
\caption{
The ACF of absolute returns for 2014--2016 data.
}
\end{figure}

For larger sampling periods, we might expect that the return distribution approaches a Gaussian distribution.
However, the convergence to the Gaussian distribution is very slow, as seen in Fig.4, which shows kurtosis as a function of the sampling period $\Delta t$.
Fig.4 indicates that a sampling period of at least several weeks is needed to approach a Gaussian distribution.

Fig.5 shows the skewness of returns as a function of $\Delta t$.
As seen in Fig.5 (left), we observe negative skewness at small $\Delta t$ shorter than one day.
On the other hand, at large $\Delta t$ longer than about one week, 
as in Fig.5 (right), the skewness is statistically consistent with zero within 1-sigma error.
For daily returns, we obtain skewness of -0.779(31).
The negative skewness of Bitcoin prices at time scales shorter than one day agrees with the previous results\cite{bouri2016return,bariviera2017some}.

\begin{figure}
\centering
\includegraphics[height=4cm,width=6cm]{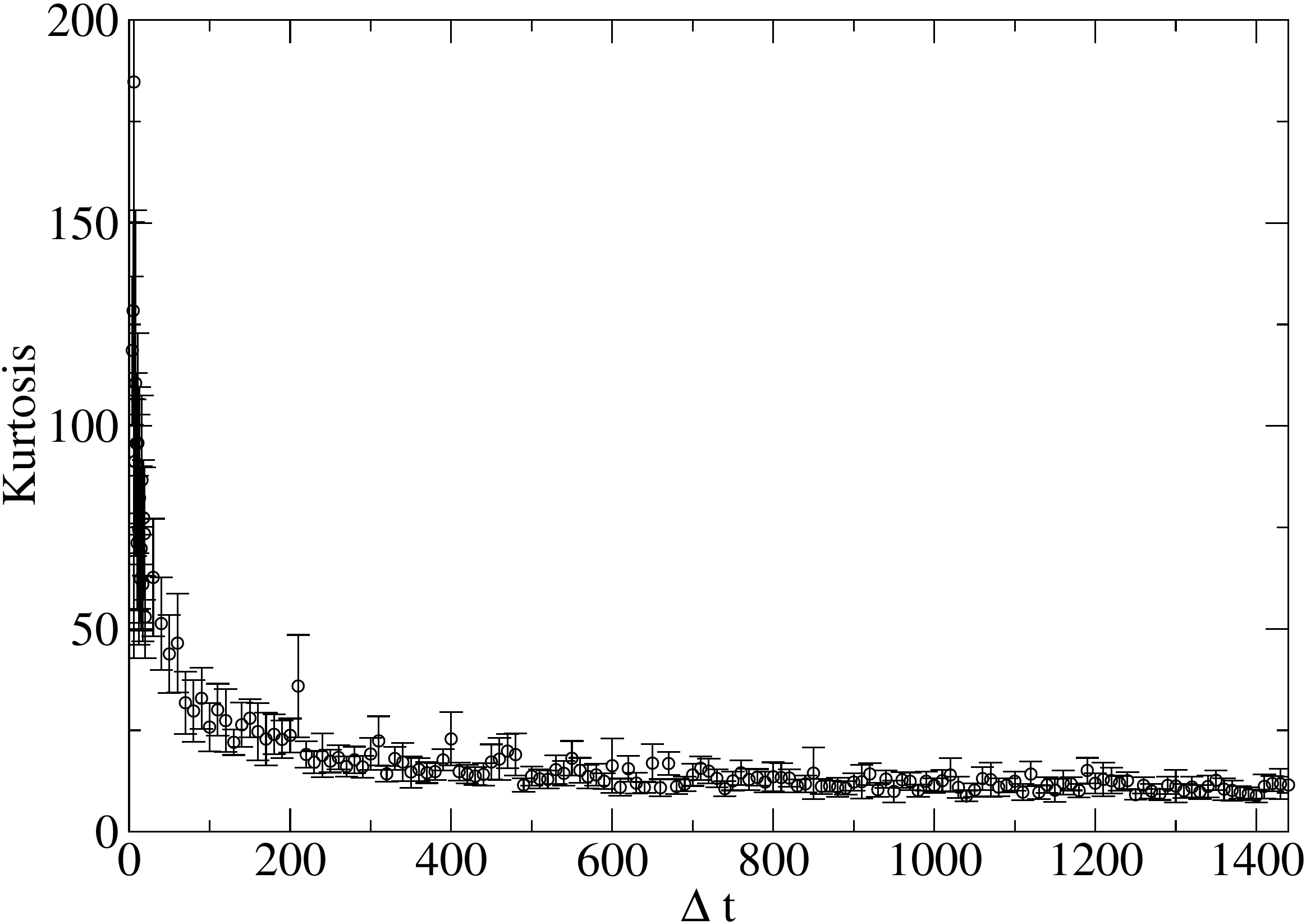}
\includegraphics[height=4cm,width=6cm]{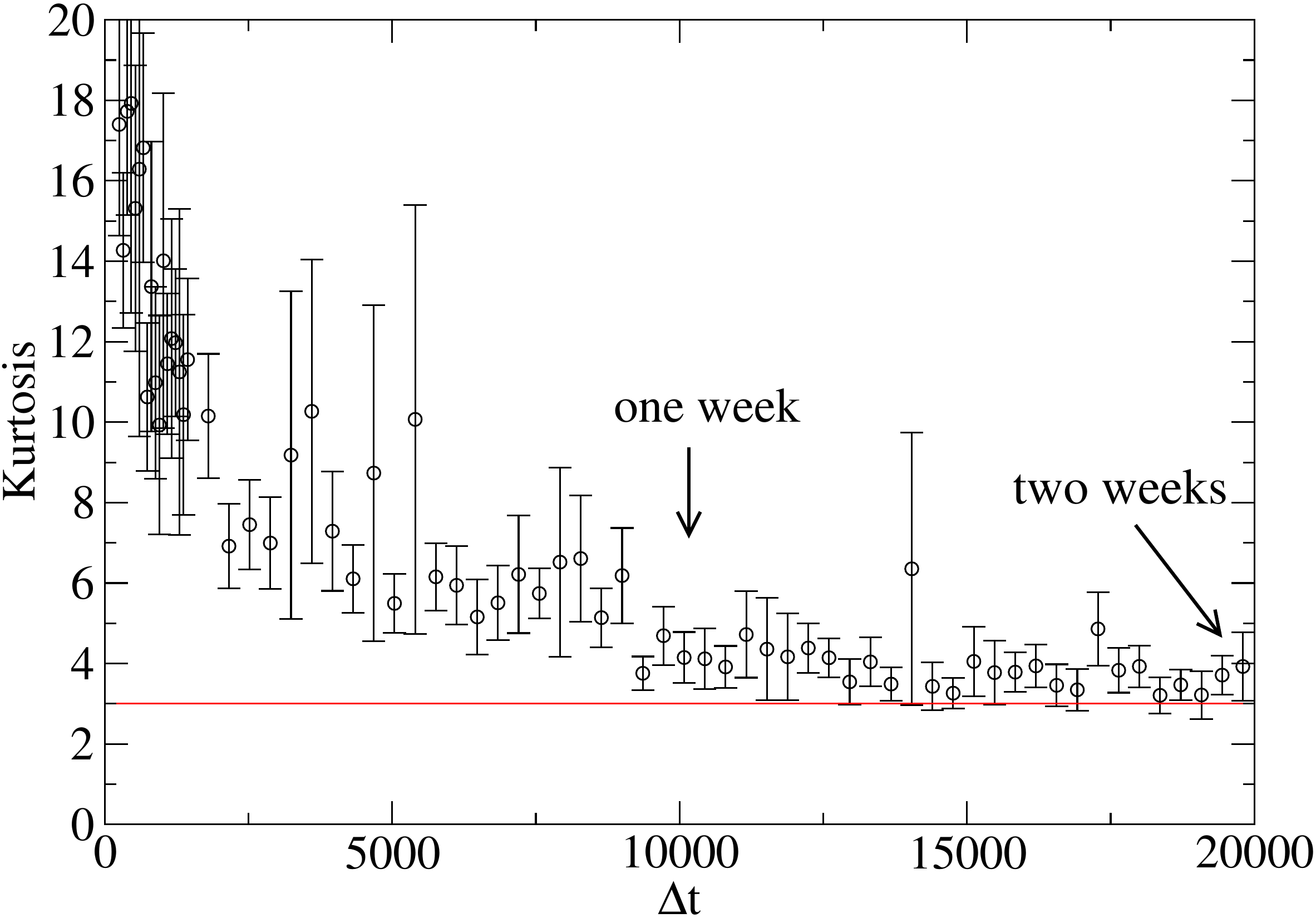}
\caption{
Kurtosis as a function of $\Delta t$.
}

\end{figure}

\begin{figure}
\centering
\includegraphics[height=4cm,width=6cm]{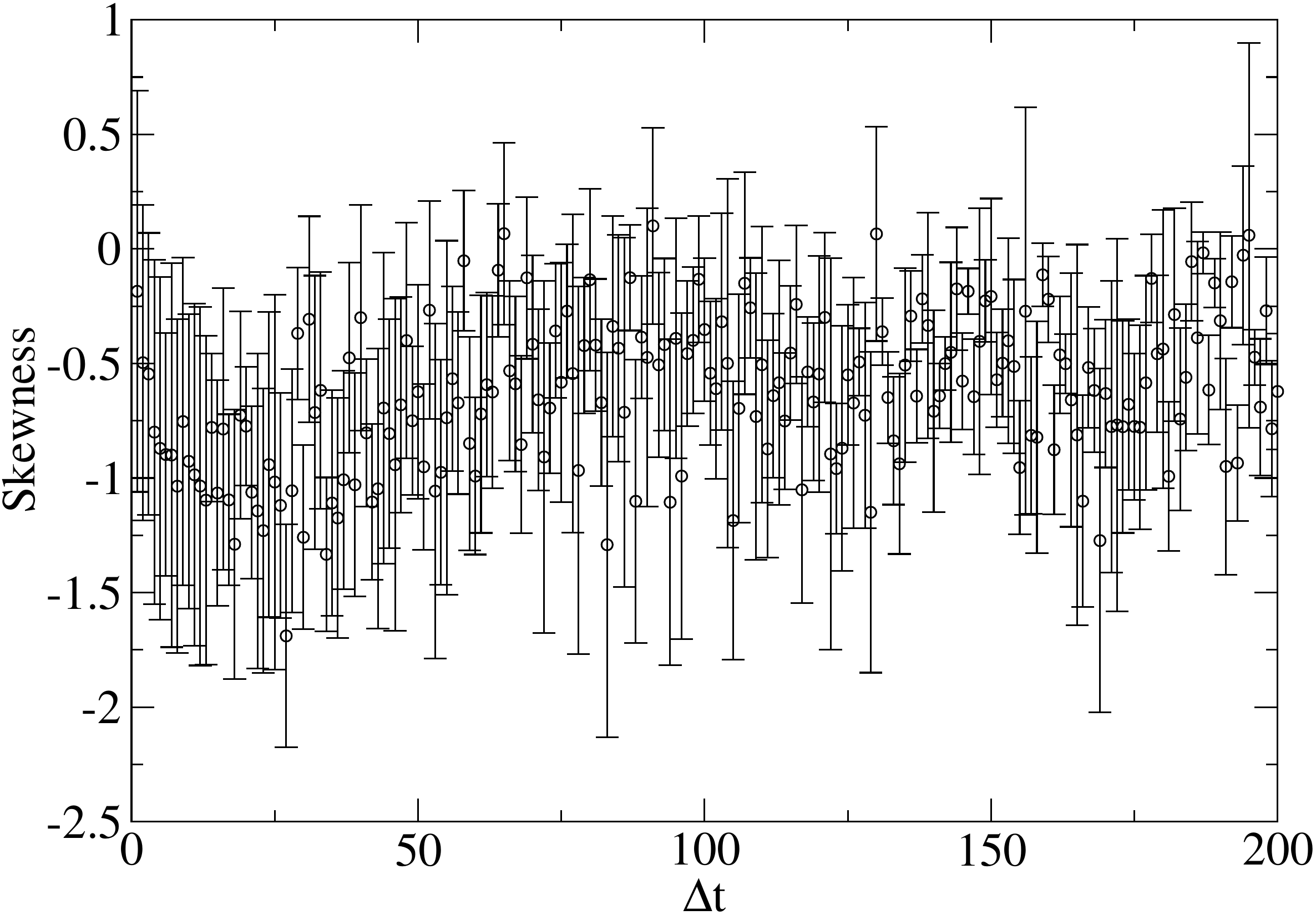}
\includegraphics[height=4cm,width=6cm]{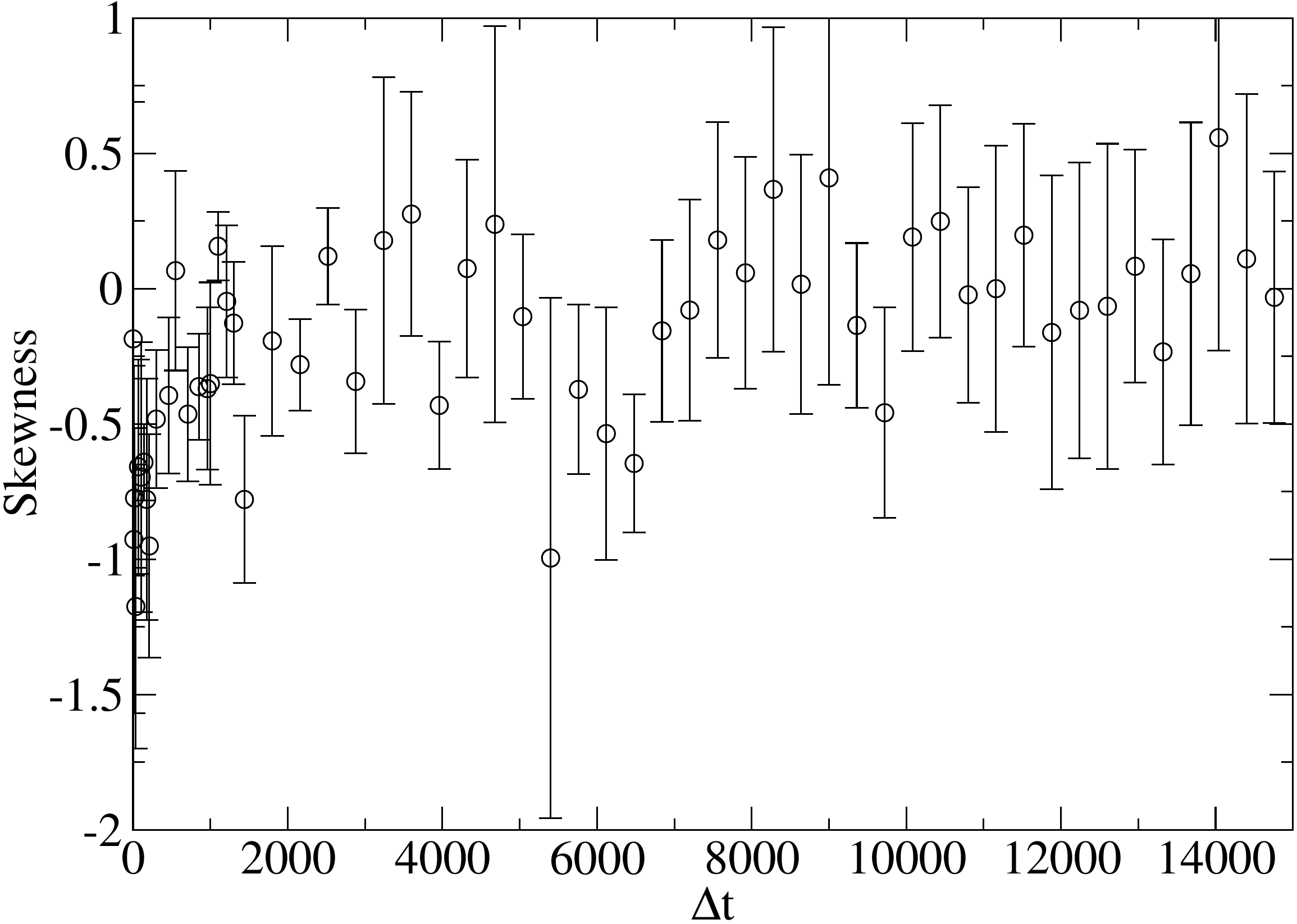}
\caption{
Skewness as a function of $\Delta t$.
}
\end{figure}

To investigate the persistence and asymmetry of volatility,
we make a volatility inference using several volatility models.
It is commonly observed that volatility is clustered and persistent for various assets.
Furthermore, for certain assets, 
the volatility process shows a higher response to negative returns.
This property of volatility is called volatility asymmetry.
Volatility asymmetry is especially common for stock price returns.
In this study, we use the generalized autoregressive conditional heteroskedasticity (GARCH)\cite{Bollerslev1986JOE},
 GJR--GARCH\cite{Glosten1993JOF} and Rational GARCH (RGARCH)\cite{takaishi2017rational} models,
for which volatility processes $\sigma_t^2$ at time $t$ are given as follows.
\begin{itemize}
\item GARCH model
\be
\sigma_t^2 = \omega + \alpha r_{t-1}^2 +  \beta \sigma_{t-1}^2,
\ee

\item GJR--GARCH model
\be
\sigma^2_t = \left\{
\begin{array}{ll}
\omega + \alpha r_{t-1}^2 + \beta \sigma_{t-1}^2,   &  r_{t-1} \ge 0,\\
\omega + (\alpha+\delta) r_{t-1}^2 + \beta \sigma_{t-1}^2,   &  r_{t-1} < 0,
\end{array}
\right.
\ee

\item RGARCH model
\be
\sigma_t^2 =\frac{\omega + \alpha r_{t-1}^2 +  \beta \sigma_{t-1}^2 }{1+\gamma r_{t-1}},
\ee
\end{itemize}
where the daily return $r_t$ is defined by $r_t=\sigma_t \epsilon_t$ and $\epsilon_t \sim N(0,1)$.
The daily return data are given by  
$r_t=100\times [log P(t)-log P(t-\Delta t)]$ with $\Delta t=1440$-min.
The model parameters to be estimated are $\alpha,\beta,\omega,\delta$ and $\gamma$.
The GARCH model is a symmetric volatility model, while the other two are asymmetric volatility models, where
the parameters of $\delta$ and $\gamma$ measure the strength of the asymmetry.
For parameter estimations, we use Bayesian inference performed by the Markov Chain Monte Carlo simulations\cite{Takaishi2009Complex,Takaishi2009,takaishi2010bayesian}.
Table 1 reports the results of parameter estimations, together with the Akaike information criterion (AIC) and deviance information criterion (DIC)\cite{Spiegelhalter2002}.
First, it is found that $\alpha +\beta$ is close to one, which means that volatility is persistent over time.
Second, we find that $\delta$ and $\gamma$ are very small or consistent with zero, which indicates that volatility asymmetry is small.
The small size of $\delta$ and $\gamma$ is consistent with the values of $\alpha,\beta$ and $\omega$ over the models being very similar to each other.
These findings conclude that the asymmetry of volatility is small for daily returns.
For small asymmetric volatility, all three models  will be approximately identical. 
It turns out that the results of AIC and DIC are very similar among models, 
which is consistent with the fact that the three models are approximately identical. 
Previous results\cite{bouri2016return} also provide no evidence of asymmetric volatility
except for the period before the price crash observed in 2013. 
Moreover, the analysis of the exponential GARCH model Ref.\cite{dyhrberg2016bitcoin} also concludes 
that Bitcoin time series does not have  an asymmetric impact on the volatility of returns.

\begin{table}
\centering
\caption{
Results of parameters. The values in parenthesis show the standard deviation.}
\label{t4}
\hspace{-10mm}
\begin{tabular}{lcccccc}
\hline
                  & GARCH  & GJR & RGARCH      \\
\hline
$\alpha$          & 0.114(18) & 0.116(19) & 0.116(19)  \\
$\beta$           & 0.878(19) & 0.874(20)  & 0.875(20)  \\
$\omega$          & 0.239(78) & 0.247(82)  & 0.249(80)  \\
$\delta$          &  ---         & 0.008(18)  &    ---        \\
$\gamma$          &   ---        &    ---        & 0.0053(47) \\
\hline
AIC              &   5539.50     &  5537.97        & 5541.55     \\
DIC              &   5535.68     &  5542.94        & 5536.73     \\
\hline
\end{tabular}
\end{table}

\section{Multifractal Analysis}

Following the MF--DFA in Ref.\cite{kantelhardt2002multifractal}, we investigate multifractality of the Bitcoin time series.
The MF--DFA is described as follows.

(i) Determine the profile $Y(i)$,
\be
Y(i)=\sum_{j=1}^i (r(j)- \bra r \ket),
\ee
where $\bra r \ket$ stands for the average of returns.

(ii) Divide the profile $Y(i)$ into $N_s$ non-overlapping segments of equal length $s$, where $N_s \equiv {int} (N/s)$.
Since the length of the time series is not always a multiple of $s$, a short time period at the end of the profile may remain.
To utilize this part, the same procedure is repeated starting from the end of the profile.
Therefore, in total $2N_s$ segments are obtained.

(iii) Calculate the variance
\be
F^2(\nu,s)=\frac1s\sum_{i=1}^s (Y[(\nu-1)s+i] -P_\nu (i))^2,
\ee
for each segment $\nu, \nu=1,...,N_s$ and
\be
F^2(\nu,s)=\frac1s\sum_{i=1}^s (Y[N-(\nu-N_s)s+i] -P_\nu (i))^2,
\ee
for each segment $\nu, \nu=N_s+1,...,2N_s$.
Here, $P_\nu (i)$ is the fitting polynomial to remove the local trend in segment $\nu$;
we use a cubic order polynomial.

(iv) Average over all segments and obtain the $q$th order fluctuation function
\be
F_q(s)=\left\{\frac1{2N_s} \sum_{\nu=1}^{2N_s} (F^2(\nu,s))^{q/2}\right\}^{1/q}.
\label{eq:FL}
\ee
For $q=0$, the averaging procedure in Eq.(\ref{eq:FL}) cannot be directly applied.
Instead, we employ the following logarithmic averaging procedure.
\be
F_0(s)=\exp\left[ \frac1{4N_s} \sum_{\nu=1}^{2N_s} \ln (F^2(\nu,s))\right].
\ee

(v) Determine the scaling behavior of the fluctuation function.
If the time series $r(i)$ are long-range power law correlated,
$F_q(s)$ is expected to be the following functional form for large $s$.
\be
F_q(s) \sim s^{h(q)}.
\label{eq:asympto}
\ee
The scaling exponent $h(q)$ is called the generalized Hurst exponent.
For $q=2$, $h(2)$ corresponds to the well-known Hurst exponent $H$.
If $h(2)<0.5$, the time series is anti-persist and if $h(2)>0.5$, it is persistent.
For $h(2)=0.5$, the time series becomes a random walk.
The relationship between the generalized Hurst exponent and the multifractal scaling exponent $\tau(q)$, defined
from the standard multifractal formalism, is given by\cite{kantelhardt2002multifractal}
\be
\tau(q) = qh(q)-1.
\ee
The singularity spectrum $f(\alpha)$, which is another way to characterize a multifractal time series, is defined by
\be
\alpha = h(q)+qh^\prime (q),
\ee
\be
f(\alpha)=q[\alpha -h(q)]+1,
\ee
where $\alpha$ is the H\"older exponent or singularity strength.

\section{Results}

\begin{figure}
\centering
  \includegraphics[height=4cm,width=6cm]{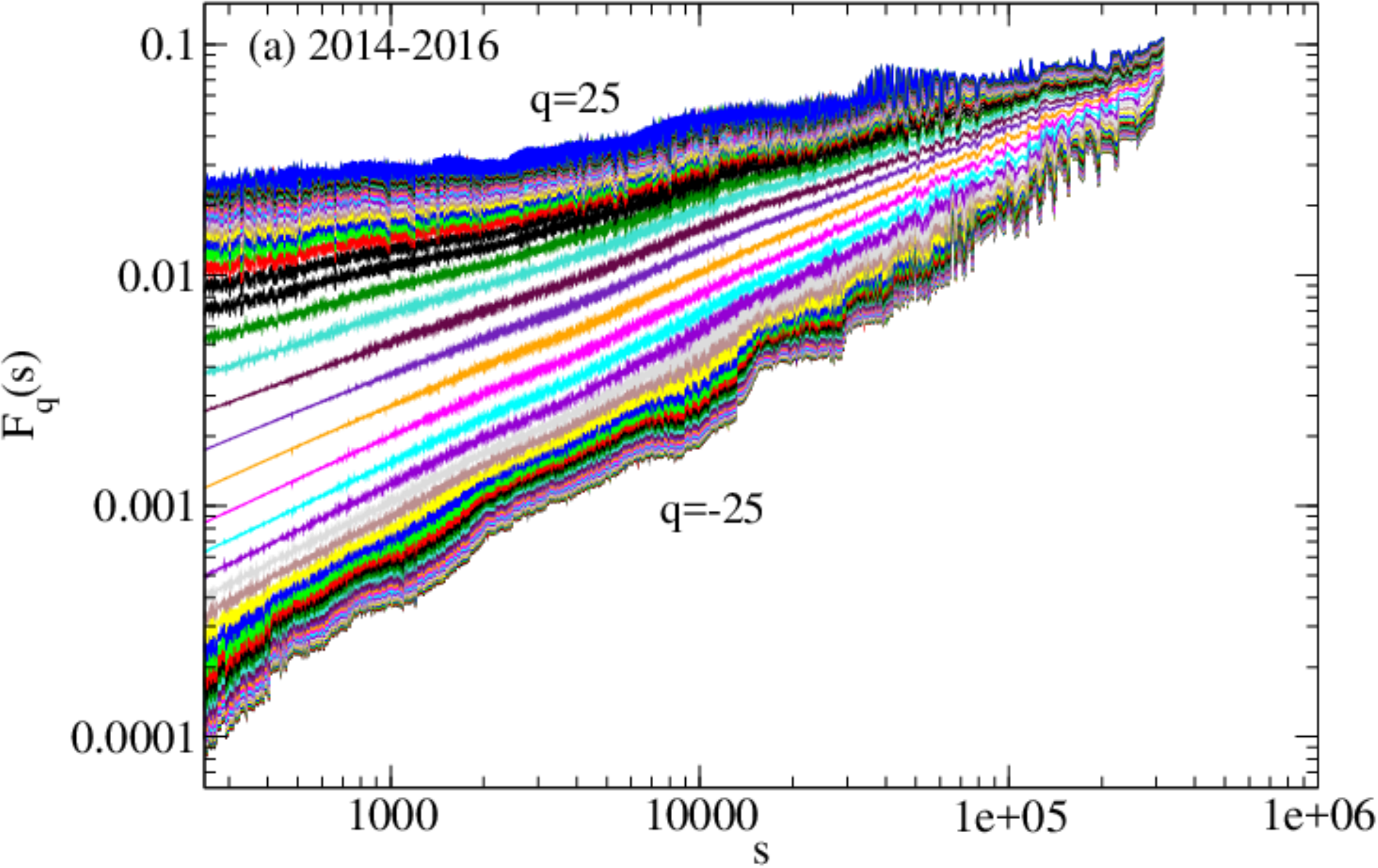}
  \includegraphics[height=4cm,width=6cm]{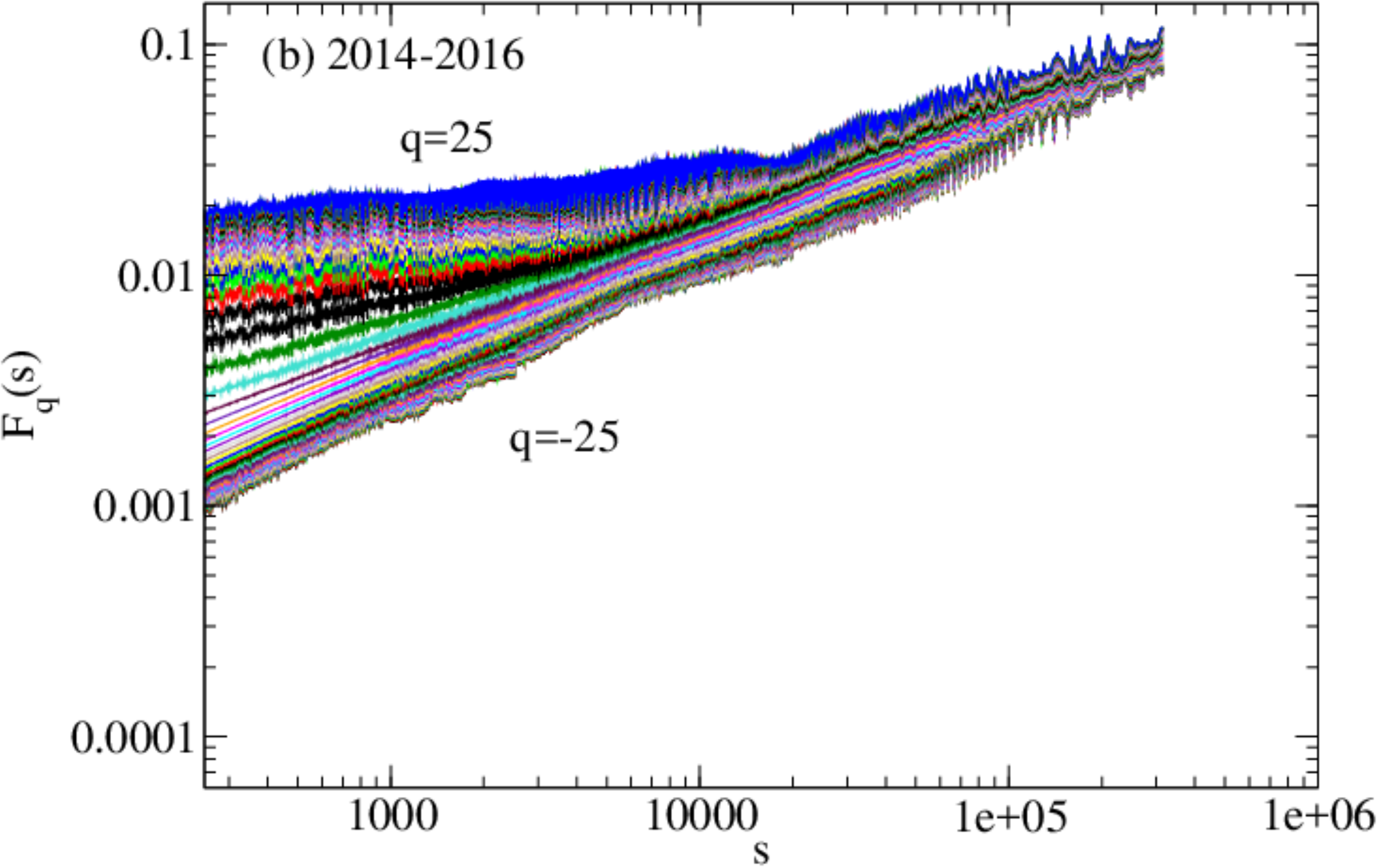}
  \includegraphics[height=4cm,width=6cm]{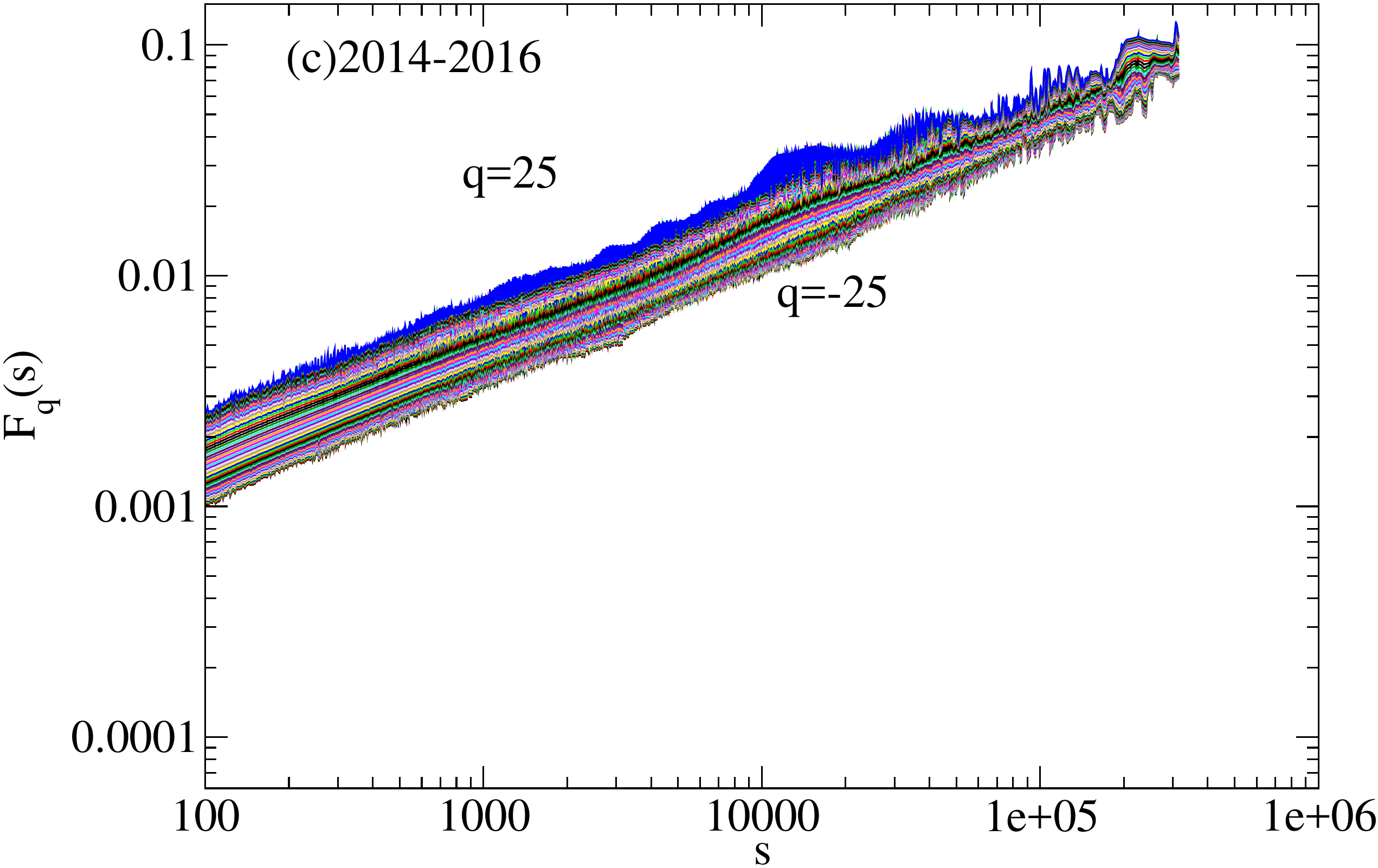}
\caption{
The fluctuation function $F_q(s)$ for (a) original data, (b) randomly shuffled data,
and (c) surrogate data in 2014--2016.
The results are plotted from $q=-25$ (bottom) to $q=25$ (up) with a step of 1.0.
}
\end{figure}

\begin{figure}
\centering
  \includegraphics[height=4cm,width=6cm]{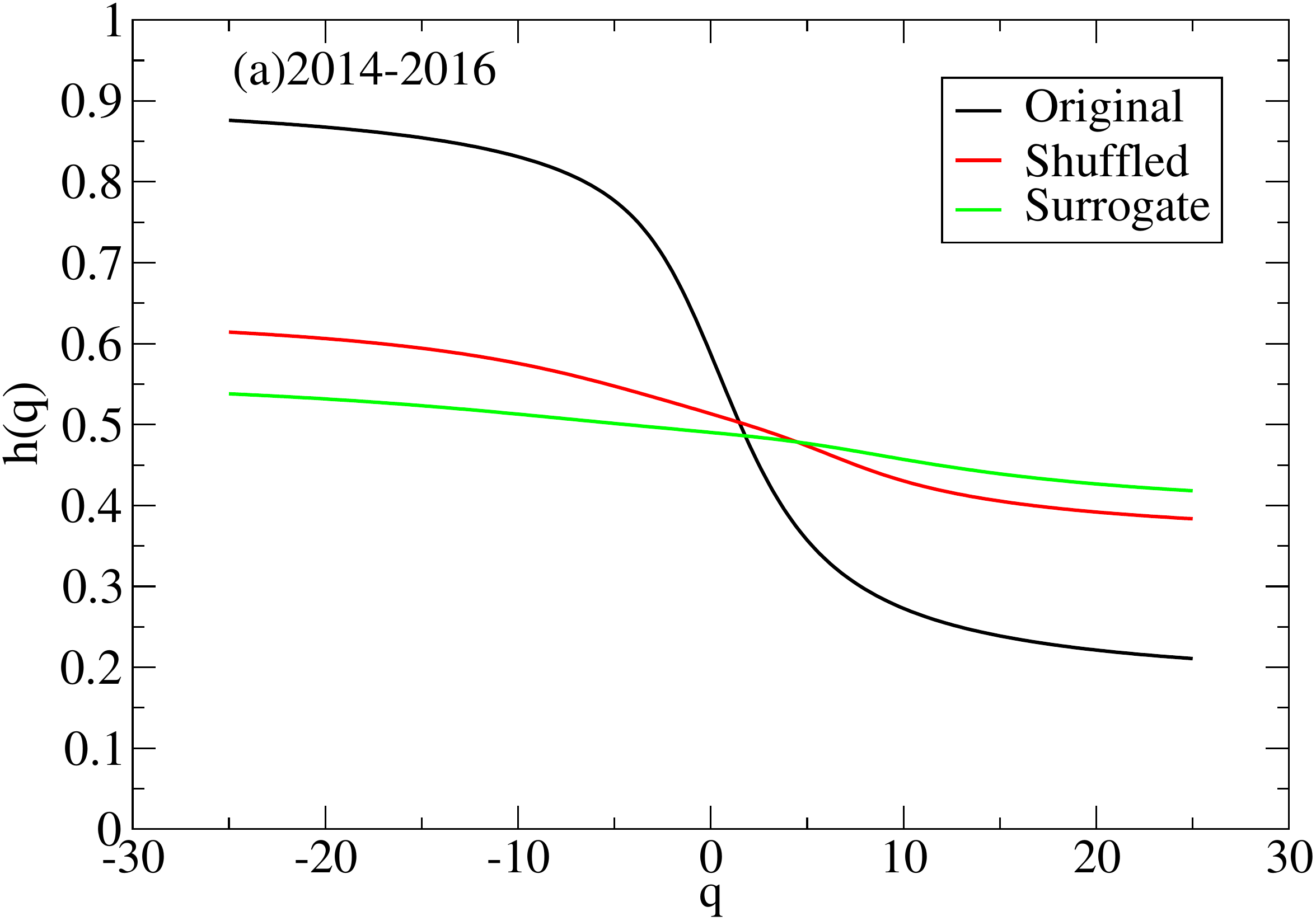}
  \includegraphics[height=4cm,width=6cm]{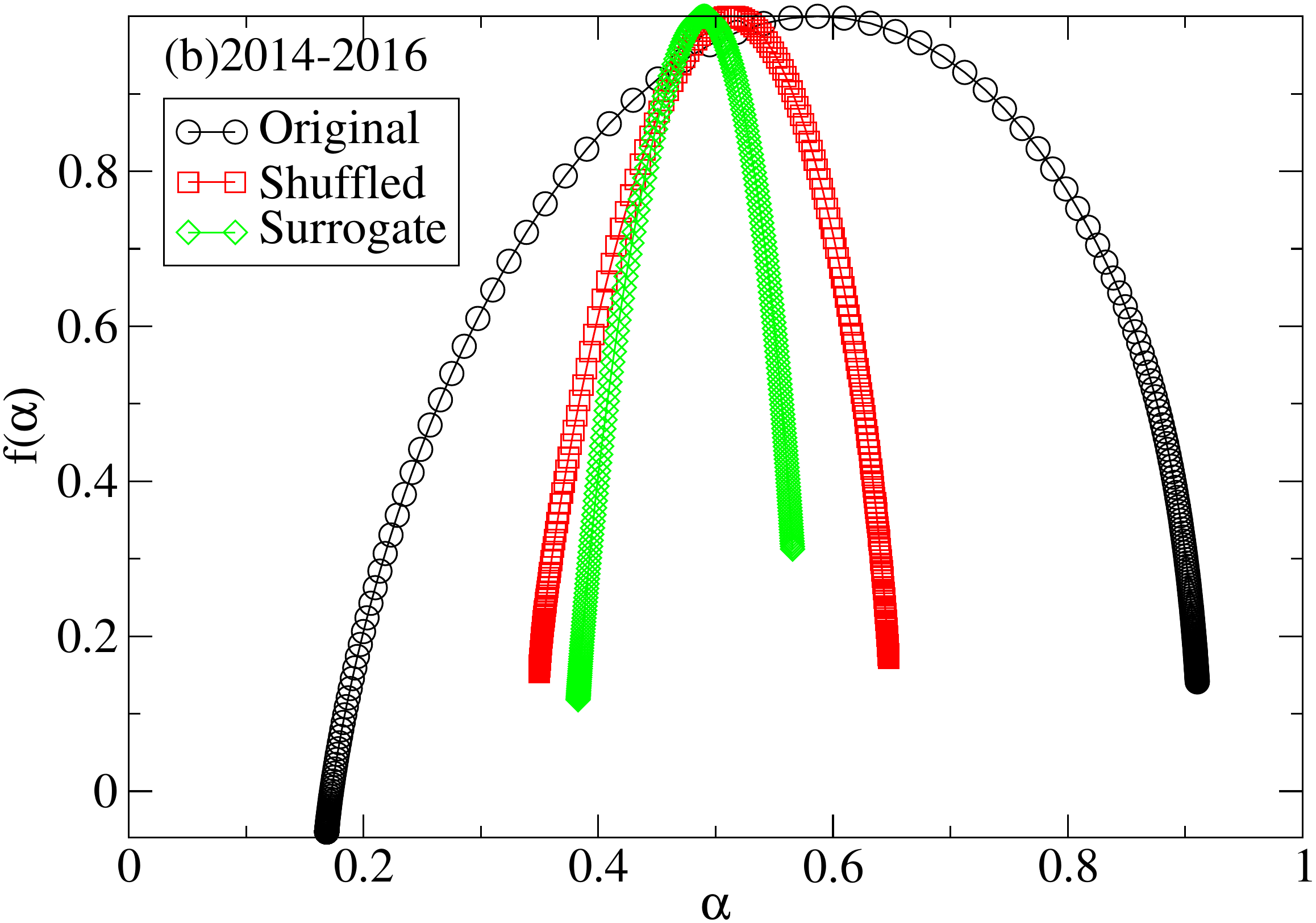}
  \includegraphics[height=4cm,width=6cm]{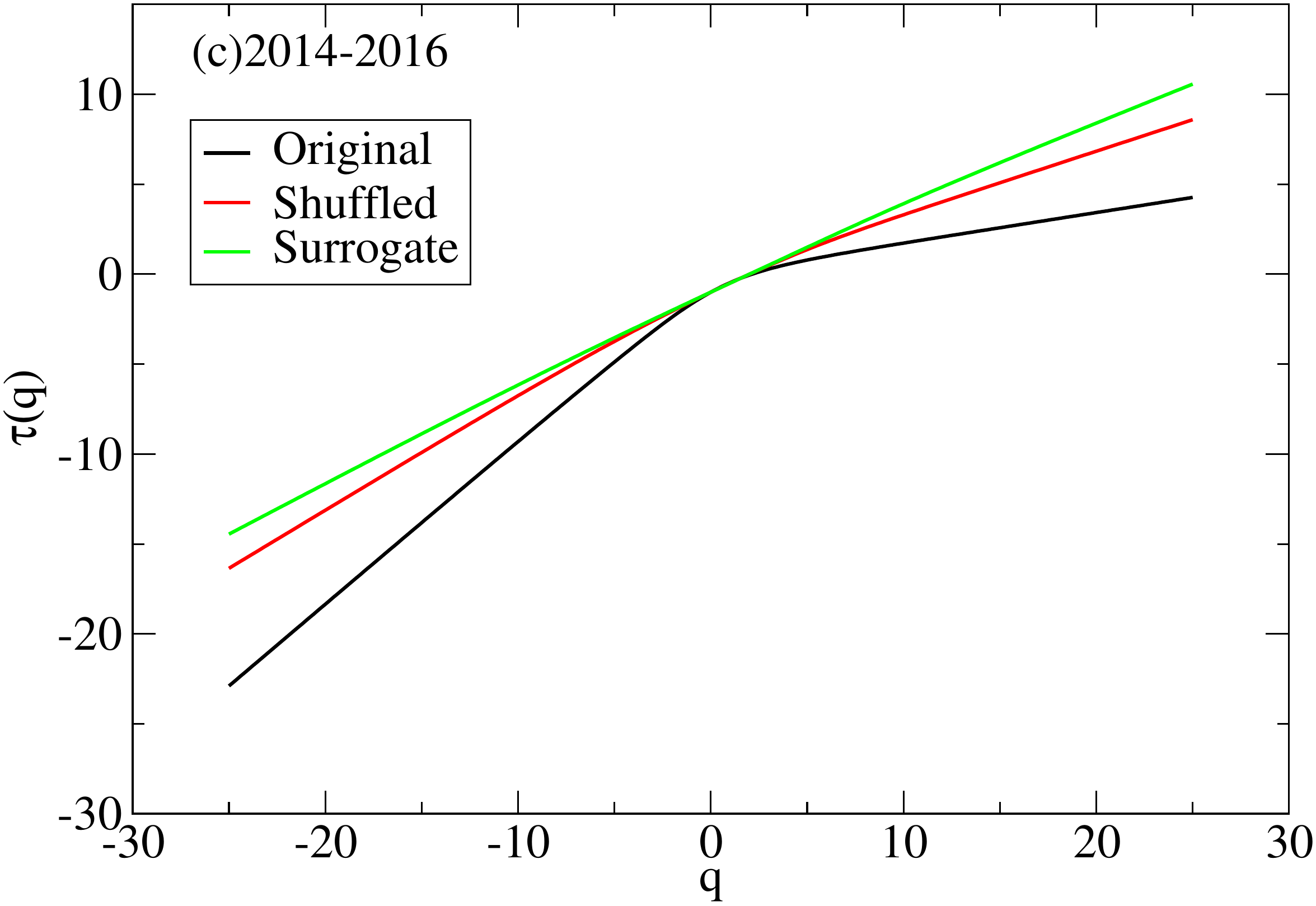}
\caption{
(a) The generalized Hurst exponent $h(q)$, (b) singularity spectrum $f(\alpha)$, 
and (c) multifractal scaling exponent $\tau(q)$ for 2014--2016.
}
\end{figure}

First, we present results analyzed using the whole dataset from January 1, 2014 to December 31, 2016.
Although we analyzed returns in several sampling periods up to $\Delta t =30$-min,
we obtained substantially the same results in each analysis. Therefore, we show only the results at $\Delta t=1$-min.
In calculating the fluctuation function $F_q(s)$, we take $q$ varying between -25 and 25, with a step of 0.2.
Fig.6(a) shows $F_q(s)$ in a log-log plot.
The slope changes depending on $q$, which indicates the multifractal property of the time series.
It has been pointed out that multifractality emerges not only because of temporal correlation,
but also because of the broad probability density\cite{kantelhardt2002multifractal}.
As seen in Fig.1, the Bitcoin return distribution turns out to be fat-tailed, and this distribution could contribute
to the multifractality of the time series.
To investigate the components of the multifractality,
we calculate the fluctuation function for the randomly shuffled time series, in which any temporal correlations have been removed,
but the same distribution has been preserved.
We also do the same calculation for the phase-randomized surrogate data. 
The phase randomization eliminates nonlinearities, preserving linear properties of the original time series data\cite{theiler1992testing}.
Fig.6(b) and (c) present the fluctuation function for the shuffled time series and surrogate data, respectively.
The figures show that the variability of the slope decreases for the shuffled time series and surrogate data.

To quantify the degree of multifractality, we calculate the generalized Hurst exponent $h(q)$, singularity spectrum $f(\alpha)$, and multifractal scaling exponent $\tau(q)$.
We extract $h(q)$ by fitting to a power law function in the range of $3000 \leq s \leq 270000$ ($100 \leq s \leq 100000$)
for the original data and shuffled data (surrogate data).
Fig.7(a), (b), and (c) show $h(q)$, $f(\alpha)$, and $\tau(q)$, respectively.
We find that, compared to the original data, 
the variability of $h(q)$, $f(\alpha)$, and $\tau(q)$ for the shuffled data decreases for the shuffled data,
which indicates that the broad probability distribution is not the only source of the multifractality of the original data.
The variability of $h(q)$, $f(\alpha)$, and $\tau(q)$ for the surrogate data is found to be much narrower than others,
which indicates that the nonlinearities of the original data also contribute to the multifractality.

Following Ref.\cite{kantelhardt2002multifractal,kwapien2005components},
we examine the strength of multifractality from the temporal correlation by decomposing the generalized Hurst exponent
into $h(q)=h_{corr}(q)+h_{sh}(q)$, where $h_{corr}(q)$ and $h_{sh}(q)$ are the correlation-induced Hurst exponent
and the Hurst exponent for shuffled time series, respectively.
We also measure the Hurst exponent for the surrogate data as $h_{su}(q)$.
We examine the degree of multifractality by the variability of $h(q)$ defined by
\be
\Delta h \equiv h(q_{min})-h(q_{max}),
\ee
and then, compare the degree of the exponents by
the ratio $R=\Delta h_{corr}/\Delta h_{sh}$.
The results of $\Delta h_{corr}$, $\Delta h_{sh}$, $\Delta h_{su}$, and $R$, together with the Hurst exponent $h(2)$, 
are presented in Table 2 ( the first column ). 
We find that the multifractal degrees of the shuffled and surrogate data, 
namely, $\Delta h_{sh}$ and $\Delta h_{su}$, are smaller than that of the original data. 
The value of $R$ turns out to be greater than one, which indicates that
the temporal correlation contributes to multifractality more than the fat-tailed distribution.
This result is similar to that obtained for high-frequency stock data\cite{kwapien2005components}.

Next, we analyze year-by-year data dividing the full sample into three periods: 2014, 2015, and 2016.
The fluctuation functions of the original data and shuffled data (surrogate data) 
are fitted to a power law function in $3000 \leq s \leq 90000$ ($100 \leq s \leq 20000$).
Fig.8--10 show $h(q)$, $f(\alpha)$, and $\tau(q)$ for the 2014, 2015, and 2016 periods, respectively,
and we list the multifractal degrees, $R$ and $h(2)$, in Table 2.
The variabilities of $h(q)$ and $f(\alpha)$ for 2014 are larger than those for 2015 and 2016.
A similar observation is found in $\Delta h$ and $\Delta h_{corr}$; 
namely, $\Delta h$ and $\Delta h_{corr}$ for 2014 are larger than those for 2015 and 2016,
which indicates that the degree of multifractality for 2014 is stronger than those for 2015 and 2016.
Moreover, the temporal correlation in 2014 contributes to the multifractality more than those in 2015 and 2016.

\begin{figure}
\centering
  \includegraphics[height=4cm,width=6cm]{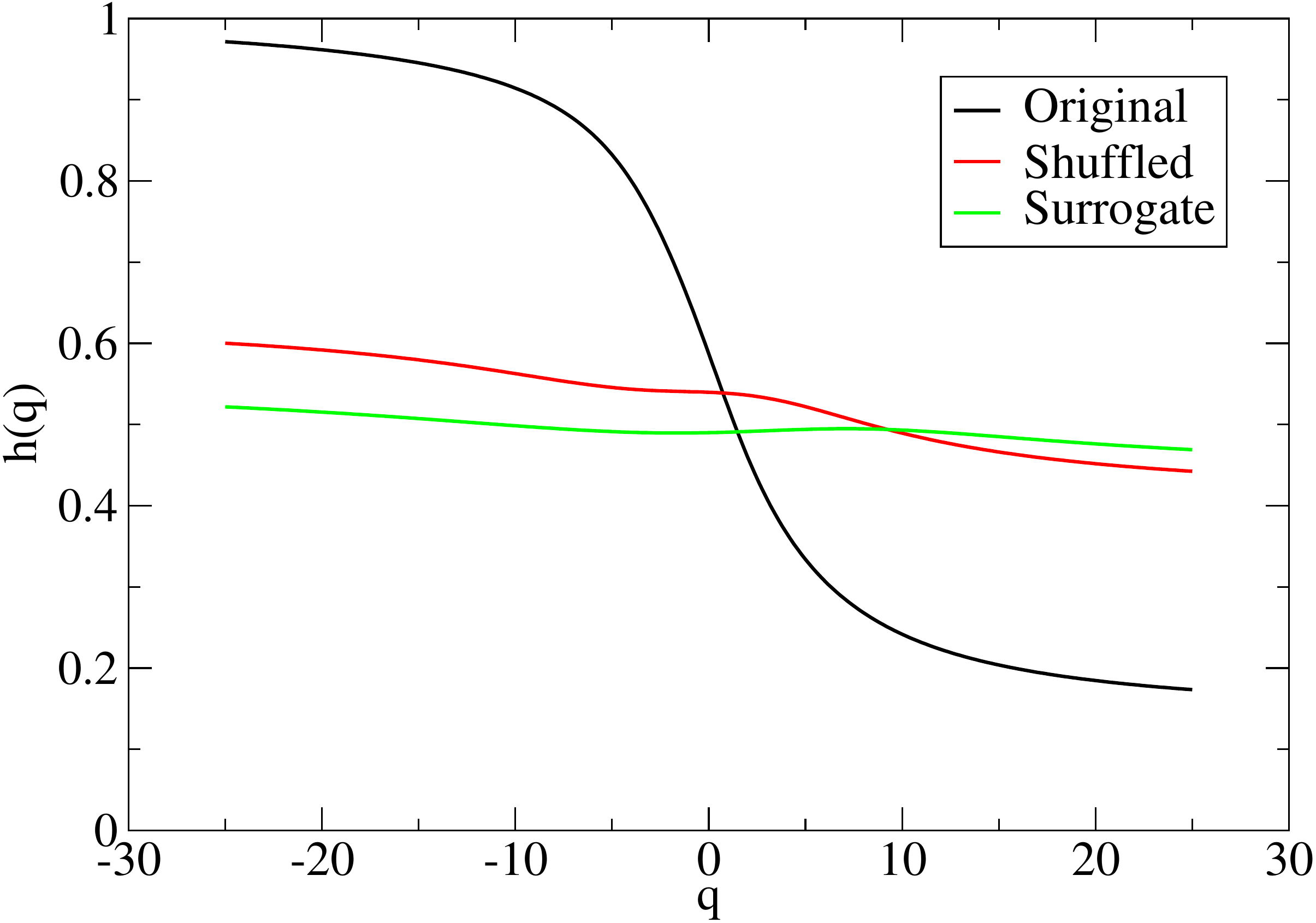}
  \includegraphics[height=4cm,width=6cm]{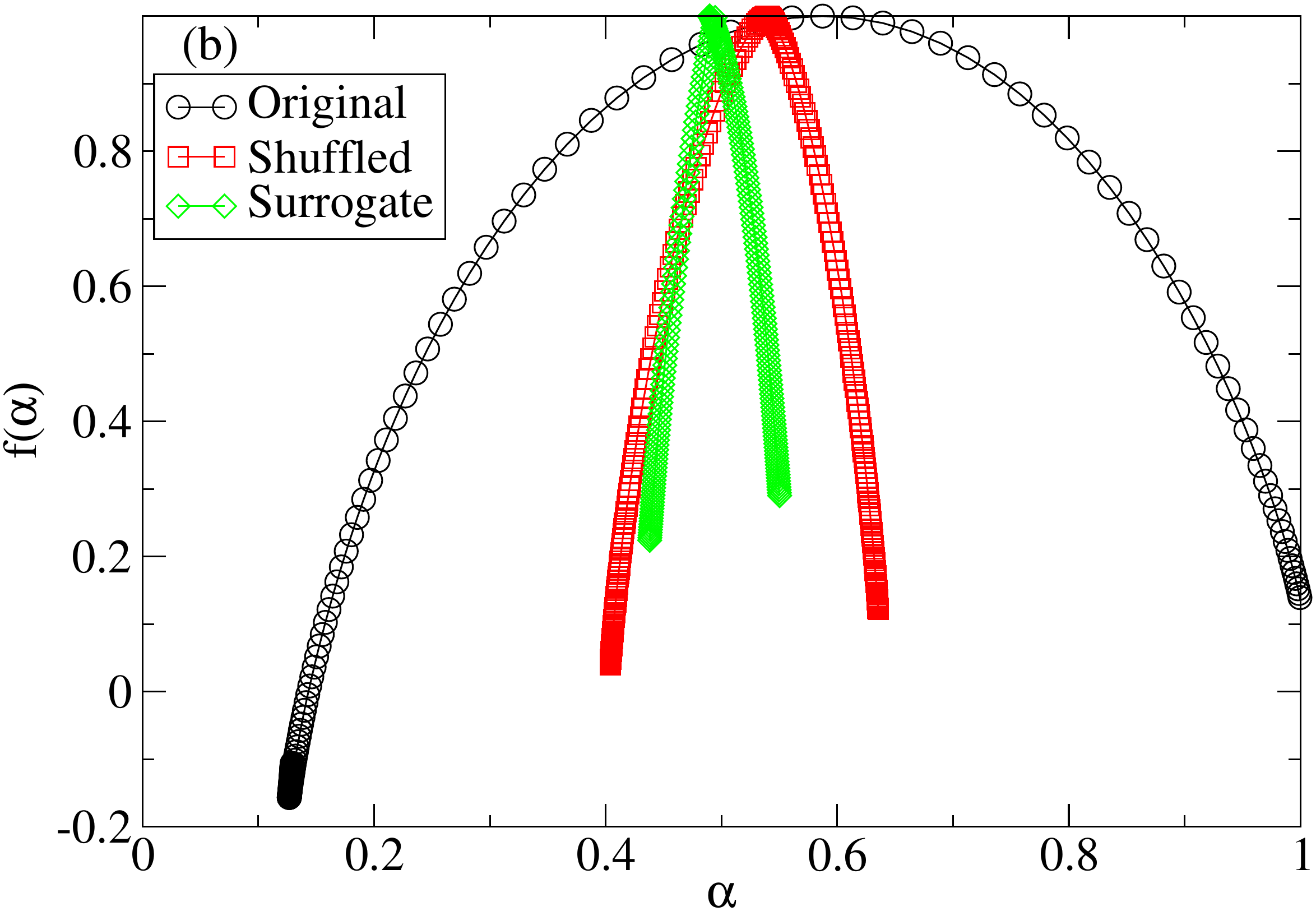}
  \includegraphics[height=4cm,width=6cm]{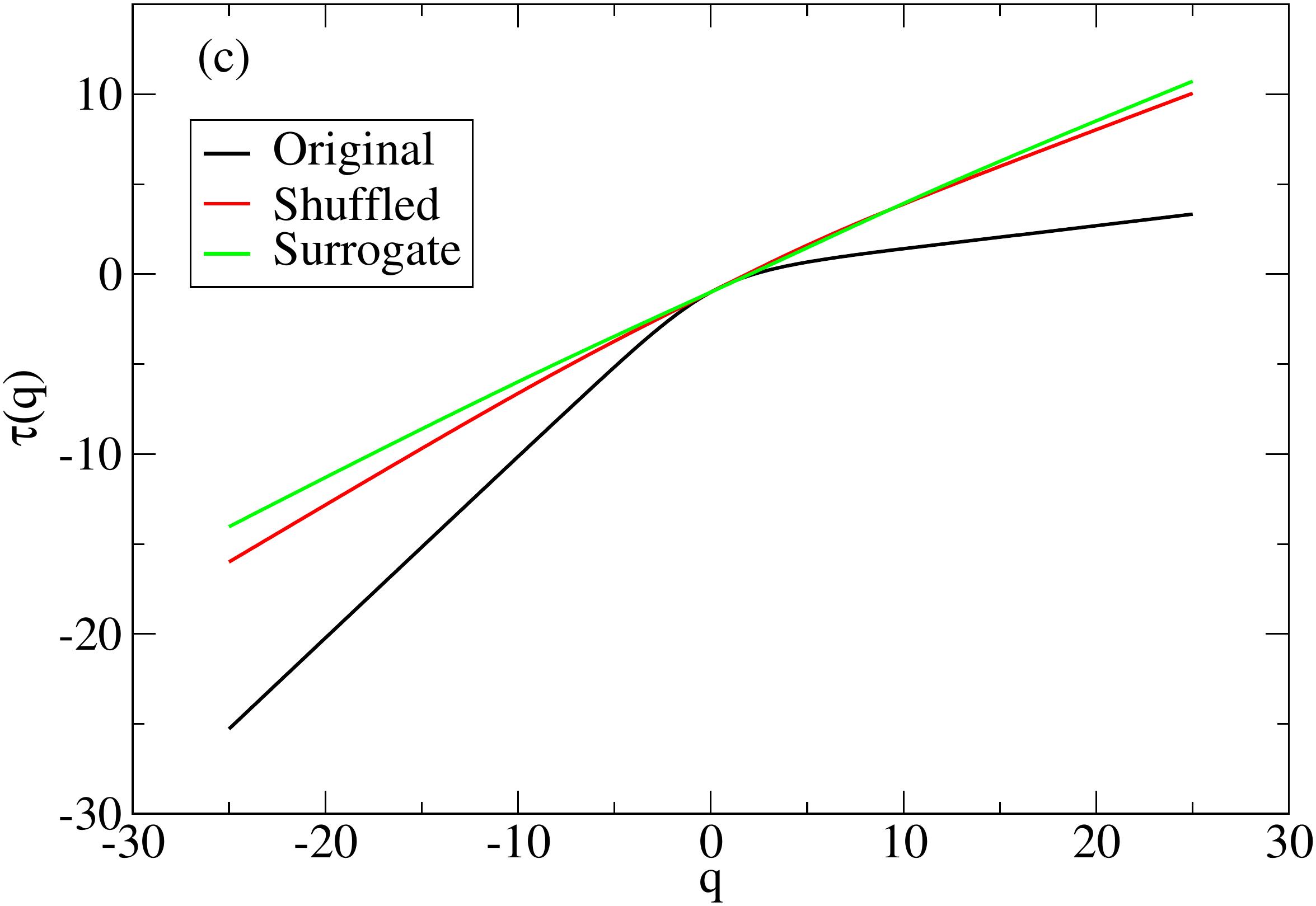}
\caption{
(a)The generalized Hurst exponent $h(q)$, (b) singularity spectrum $f(\alpha)$, and 
(c) multifractal spectrum $\tau(q)$ for 2014.
}
\end{figure}

\begin{figure}
\centering
  \includegraphics[height=4cm,width=6cm]{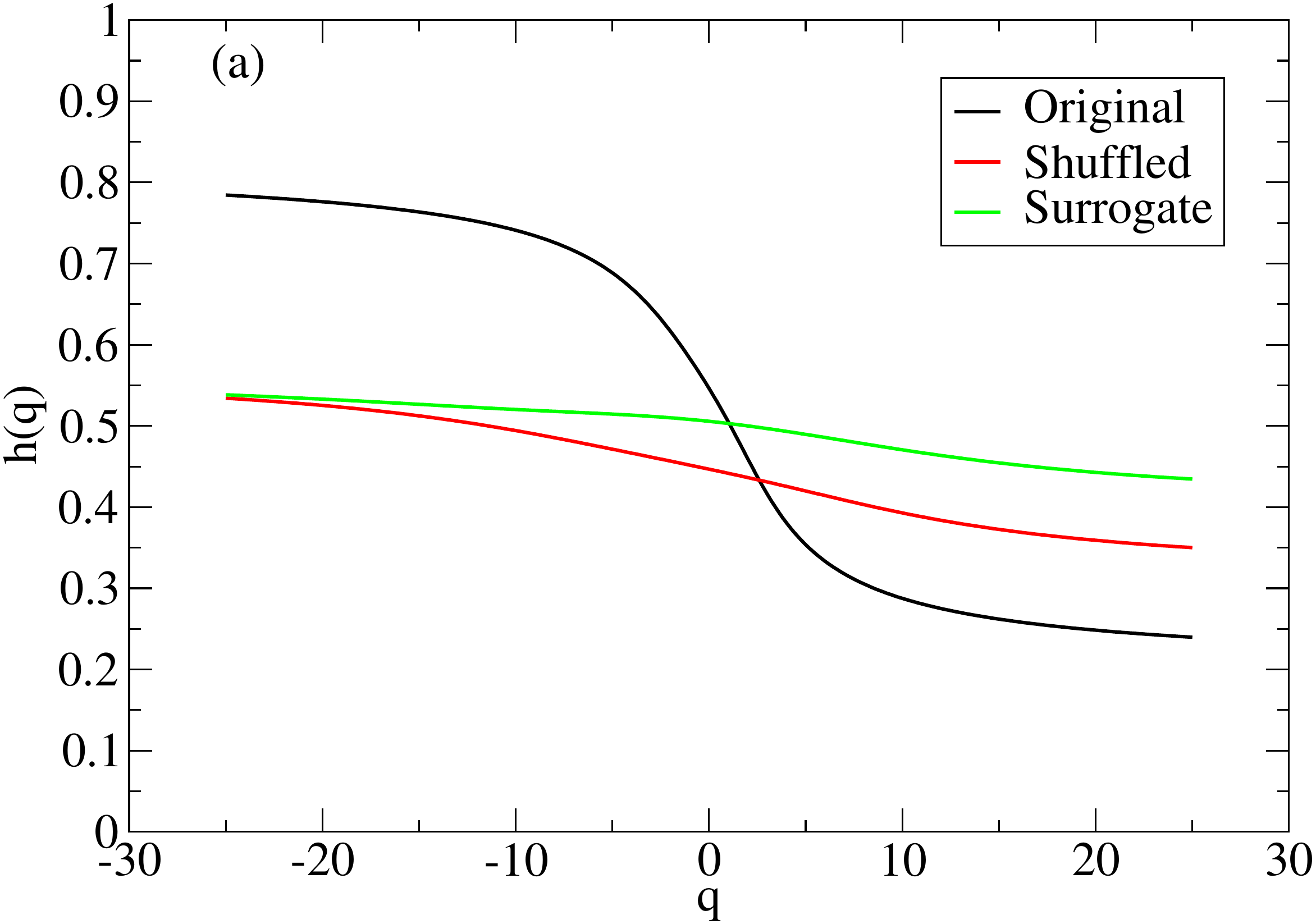}
  \includegraphics[height=4cm,width=6cm]{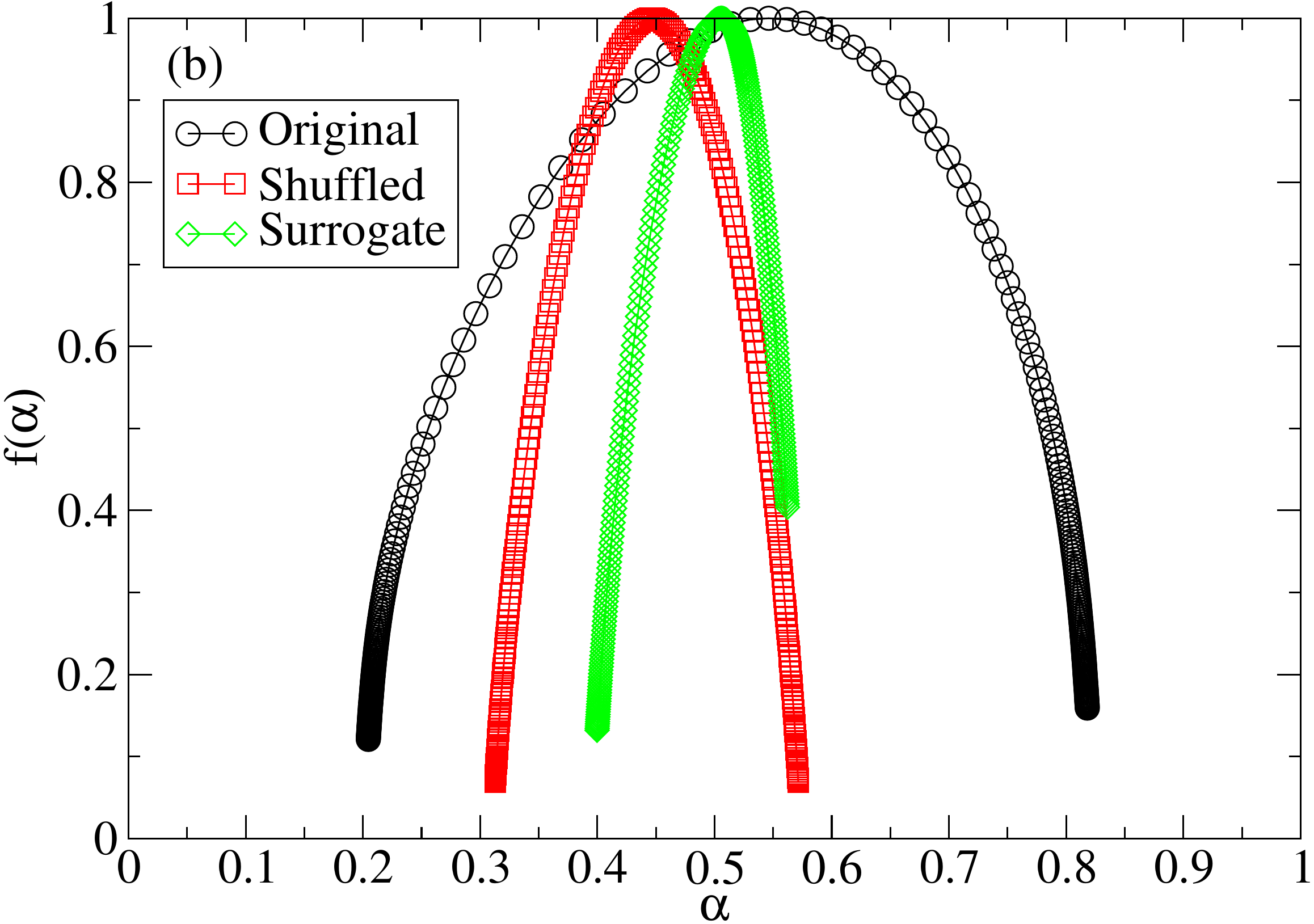}
  \includegraphics[height=4cm,width=6cm]{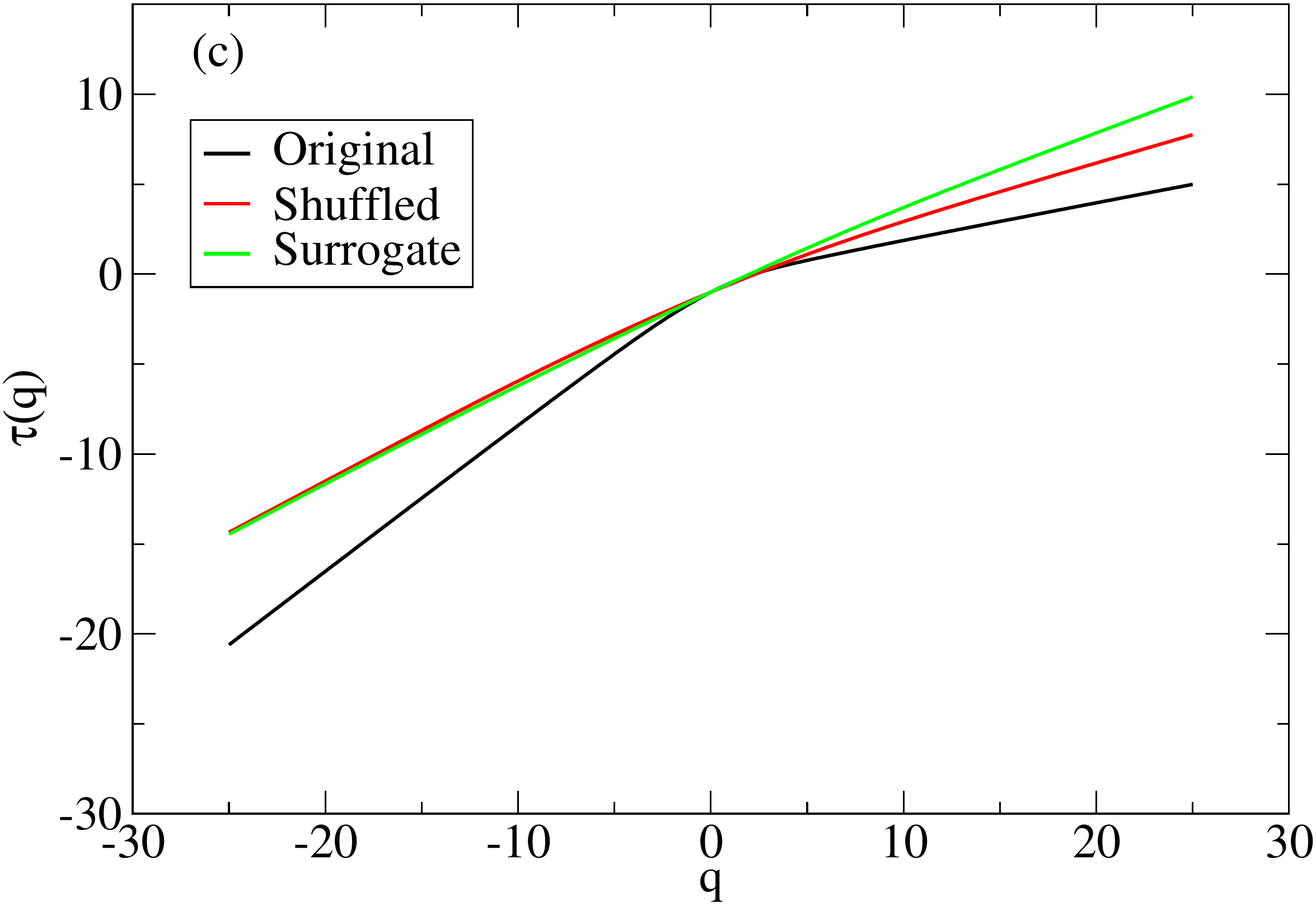}
\caption{
(a)The generalized Hurst exponent $h(q)$, (b) singularity spectrum $f(\alpha)$, and
(c) multifractal spectrum $\tau(q)$ for 2015.
}
\end{figure}

\begin{figure}
\centering
  \includegraphics[height=4cm,width=6cm]{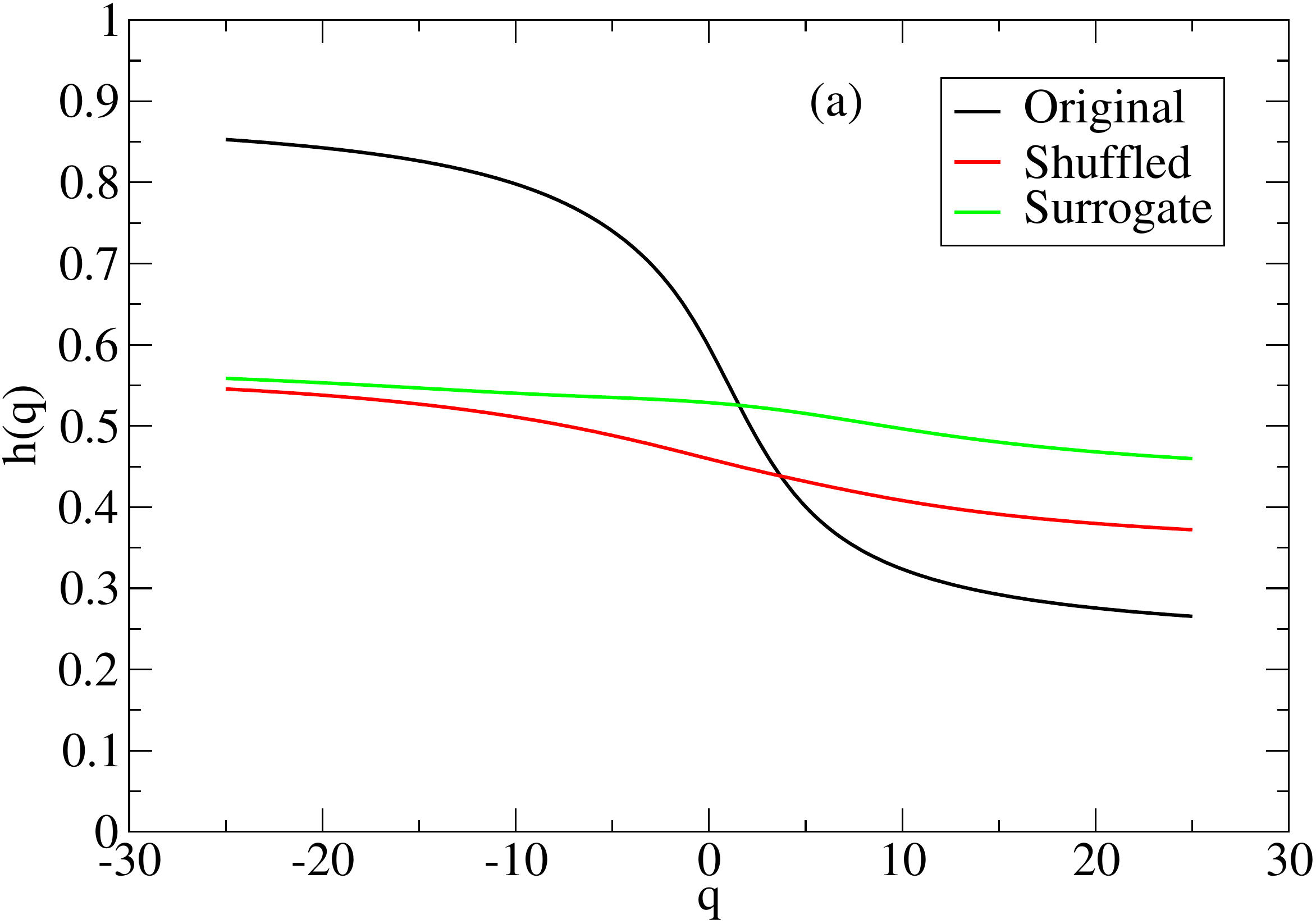}
  \includegraphics[height=4cm,width=6cm]{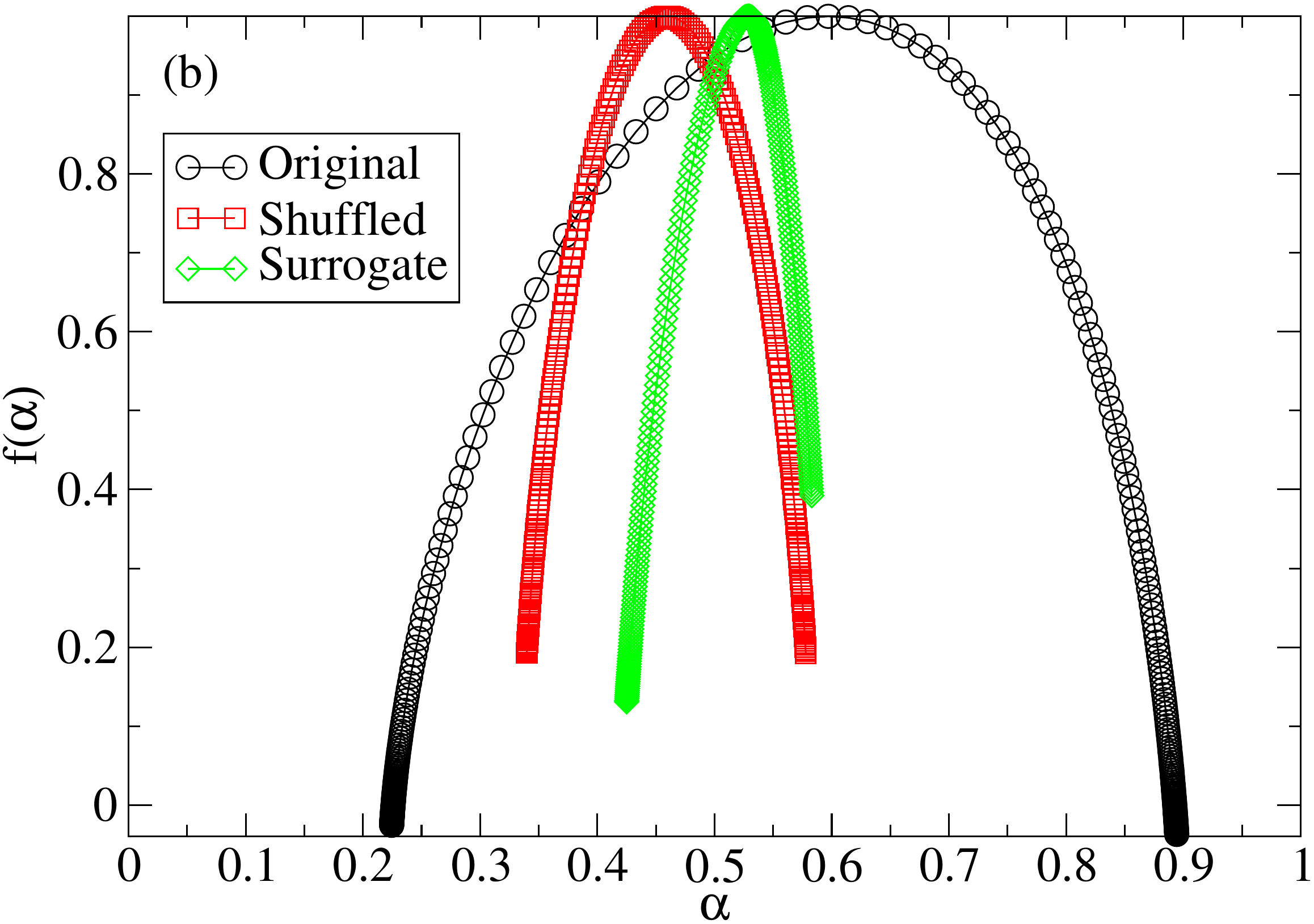}
  \includegraphics[height=4cm,width=6cm]{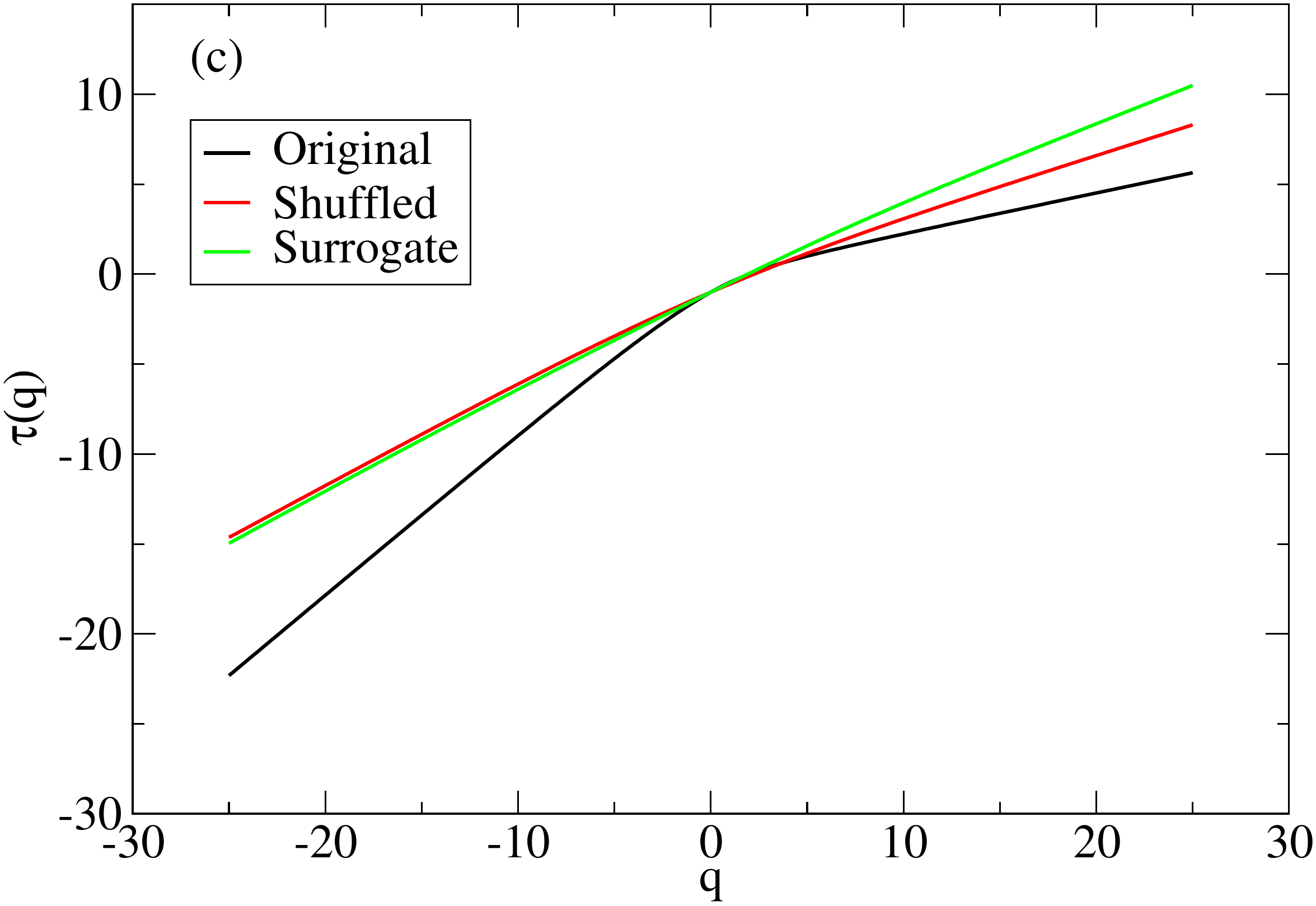}
\caption{
(a)The generalized Hurst exponent $h(q)$, (b) singularity spectrum $f(\alpha)$, and
(c) multifractal spectrum $\tau(q)$ for 2016.
}
\end{figure}

\begin{table}
\centering
\caption{
Results of $h(2)$, $\Delta h$,$\Delta h_{corr}$, $\Delta h_{sh}$, $\Delta h_{su}$ and $R$.}
\hspace{-10mm}
\begin{tabular}{lcccccc}
\hline
                  & 2014-2016  & 2014   & 2015 & 2016      \\
\hline
$h(2)$            & 0.475     & 0.461 & 0.460 & 0.505  \\
$\Delta h$        & 0.665     & 0.798 &  0.545 & 0.613 \\
$\Delta h_{corr}$ & 0.434     & 0.640 & 0.360 & 0.429 \\
$\Delta h_{sh}$      & 0.231     & 0.158 & 0.184 & 0.173 \\
$\Delta h_{su}$     & 0.120     & 0.053  & 0.104 & 0.099 \\
$R$               & 1.88      & 4.06  & 1.96  & 2.47  \\
\hline
\end{tabular}
\end{table}

\section{Comparison with GBP--USD Exchange Rate}
In this section, we analyze the multifractality of GBP--USD exchange rate.
The data analyzed are the time series of 1-min GBP--USD exchange rate for the time period from 
January 3, 2016 to December 20, 2016\cite{GBP}.
We focus on the year 2016 in which a referendum on the prospective withdrawal of the United Kingdom from the European Union (EU) 
was taken on June 23, and majority voted to leave the EU. This incident is also referred to as "Brexit."
We investigate whether Brexit influences properties of GBP--USD exchange rate and Bitcoin time series. 

In Fig.11, we shows $h(q)$, $f(\alpha)$, and $\tau(q)$ for 
the original, shuffled, and surrogate data in 2016. 
The properties of $h(q)$, $f(\alpha)$, and $\tau(q)$ are similar to those of the Bitcoin time series;
namely, the variabilities of $h(q)$, $f(\alpha)$, and $\tau(q)$ decrease for the shuffled and surrogate data,
and the broad probability distribution and temporal correlations contribute to the multifractality of the original data.

\begin{figure}
\centering
  \includegraphics[height=4cm,width=6cm]{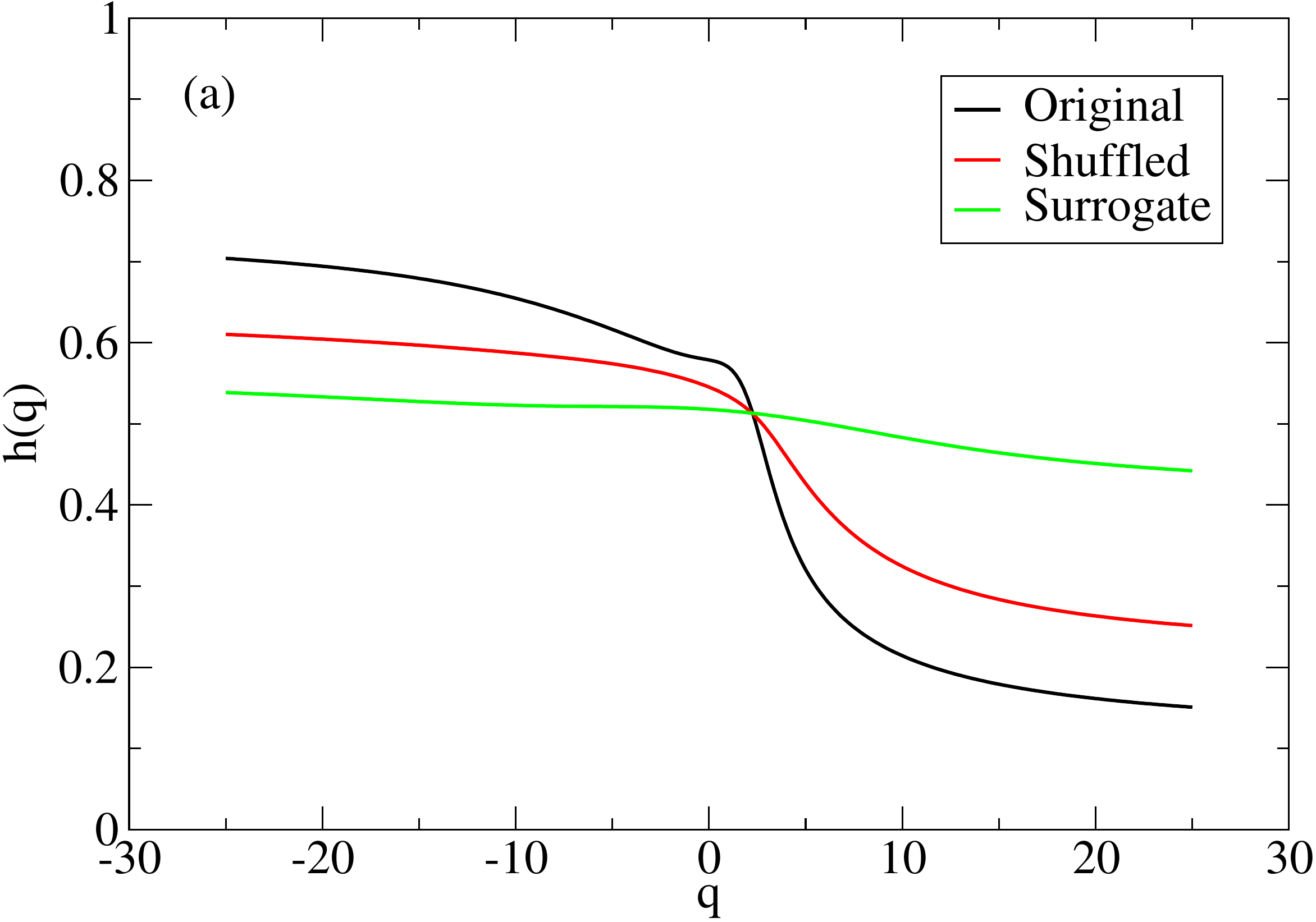}
  \includegraphics[height=4cm,width=6cm]{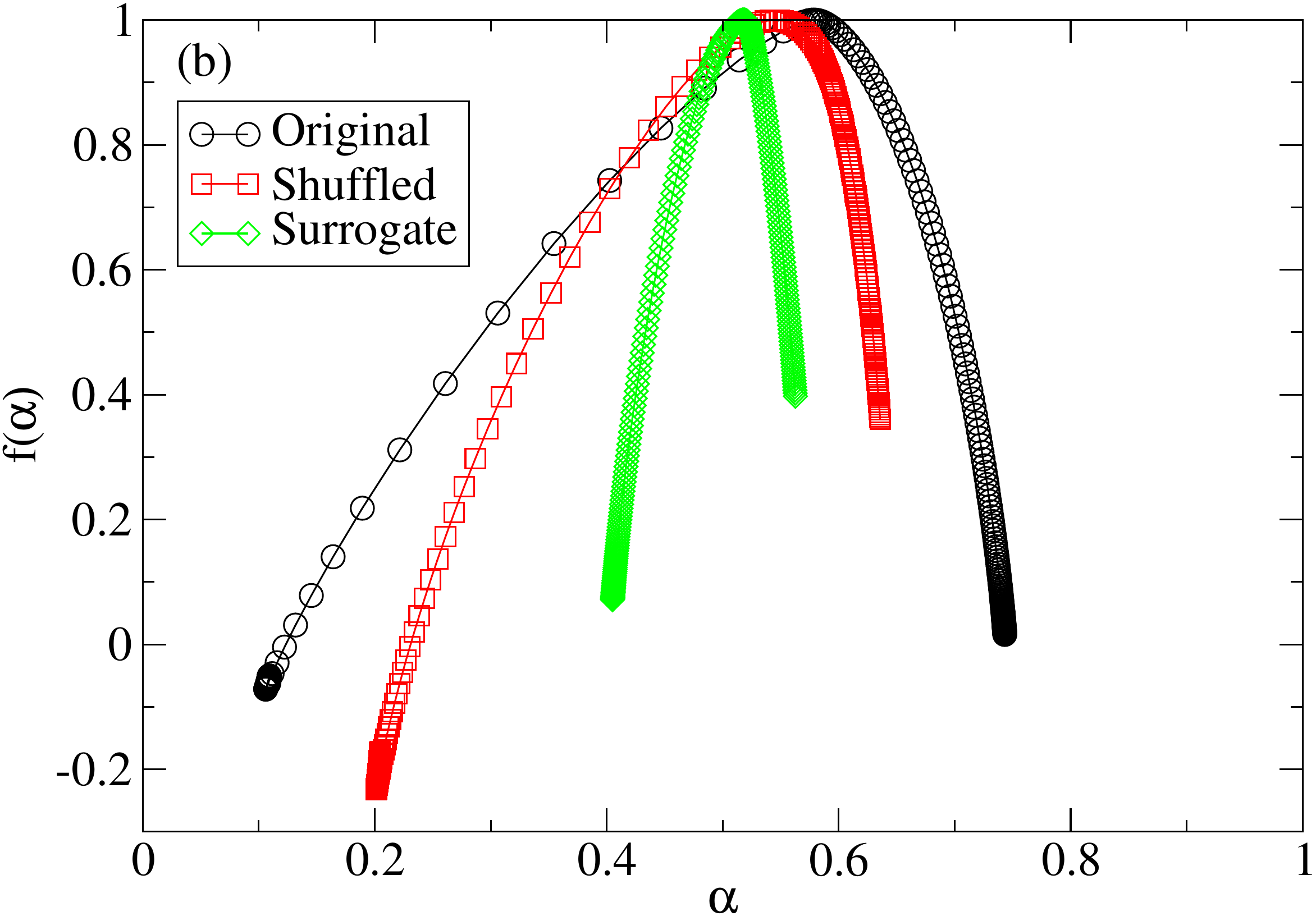}
  \includegraphics[height=4cm,width=6cm]{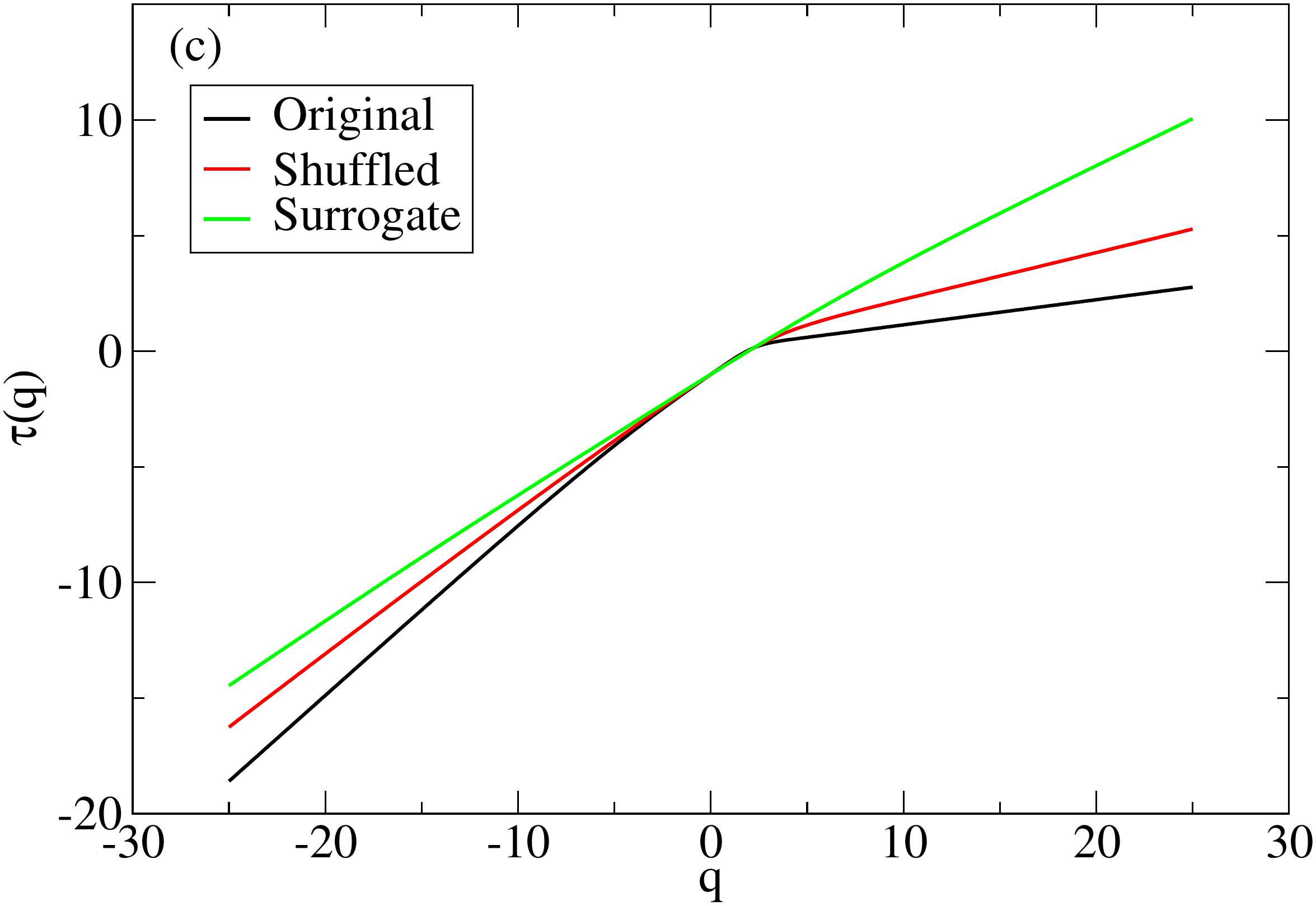}
\caption{
(a)The generalized Hurst exponent $h(q)$, (b) singularity spectrum $f(\alpha)$, and
(c) multifractal spectrum $\tau(q)$ for the GBP--USD exchange rate time series in 2016.
}
\end{figure}

To investigate the time evolution of the multifractality, 
we employ the method of rolling windows.   
We calculate the $q$th order fluctuation function with a window size of 30 days, and
then, determine the generalized Hurst exponent $h(q)$.
Fig.12(a) shows the 1-min GBP--USD exchange rate return time series, and dynamic evolution of $\Delta h$ and $h(2)$.
It is found that there are large changes in returns around "Brexit," and the multifractal degree $\Delta h$  
shows a spike-like change. The Hurst exponent $h(2)$ also changes and increases at Brexit. 
These observations indicate that Brexit does influence the GBP--USD exchange rate.

We do the same analysis for Bitcoin time series in 2016.
Fig.13 shows the 1-mine returns, dynamic evolutions of $\Delta h$ and $h(2)$ for Bitcoin time series in 2016.
While there is a volatile period in return time series around Brexit, 
we find no clear response to Brexit in $\Delta h$ and $h(2)$. 
Thus we argue that in contrast to the GBP--USD exchange rate time series,
the Bitcoin time series was robust to Brexit on June 23, 2016.

\begin{figure}
\centering
  \includegraphics[height=8cm,width=10cm]{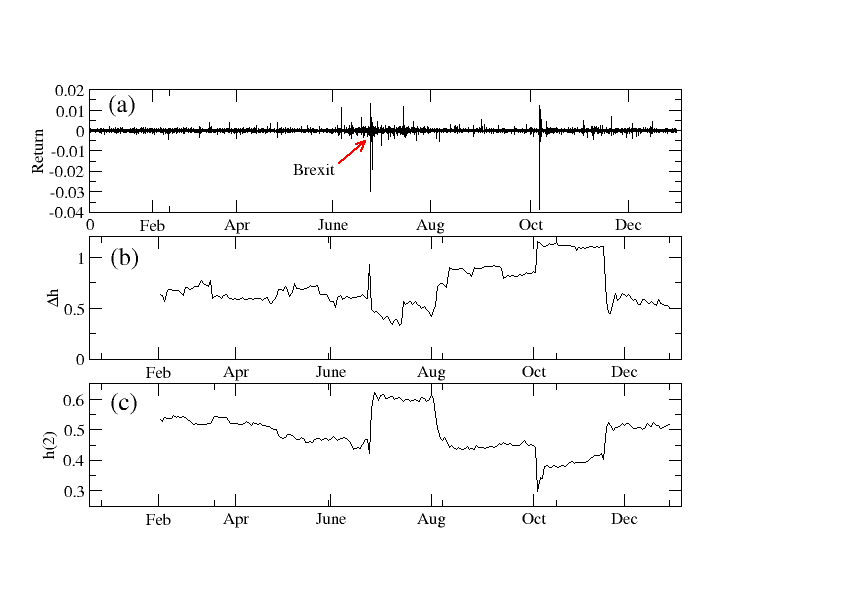}
\caption{
(a) 1-minite GBP--USD exchange rate return time series in 2016. "Brexit" indicates  the referendum on June 23, 2016.
(b) $\Delta h$. 
(c) $h(2)$.
}
\end{figure}

\begin{figure}
\centering
  \includegraphics[height=8cm,width=10cm]{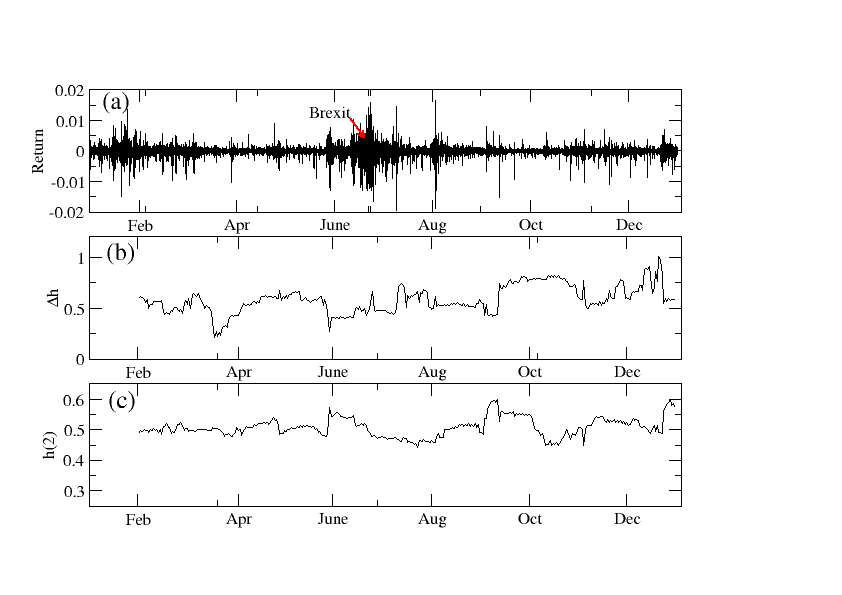}
\caption{
(a) 1-min Bitcoin return time series in 2016. "Brexit" indicates  the referendum on June 23, 2016.
(b) $\Delta h$.
(c) $h(2)$.
}
\end{figure}

\section{Conclusion}
We find that the 1-min return distribution is leptokurtic, and
the kurtosis largely deviates from the Gaussian expectation.
Calculating kurtosis as a function of the sampling period,
we find that kurtosis approaches the Gaussian expectation,
although the convergence to that is very slow. A sampling period of at least a few weeks is needed
to converge to the Gaussian expectation.
Skewness is found to be negative at time scales shorter than one day, and becomes consistent
with zero at time scales larger than about one week.
The ACF of returns is short ranged, and goes to zero at the time lag around 5-min. 
Then, the ACF overshoots zero, and becomes negative up to around 10-min. 
This overshooting phenomenon is also seen for the high-frequency returns of the DAX index\cite{VOIT2003286}. 
Contrary to returns, the ACF of absolute returns is very long ranged, and follows a power law function.
The GARCH, GJR, and RGARCH models are applied to the Bitcoin time series to
investigate the daily volatility asymmetry. 
While all the models show the volatility persistence or volatility clustering,  
we find no evidence of volatility asymmetry.
In many respects, Bitcoin exhibits the similar properties to other assets 
that are classified as the stylized facts: fat-tailed return distribution, short ranged autocorrelation in returns,
a power law function of autocorrelation function of absolute returns, and volatility clustering.

On exploring multifractality using multifractal detrended fluctuation analysis,
we find that the Bitcoin time series exhibits multifractality.
The sources of the multifractality are investigated, and
it is confirmed that both temporal correlation and the fat-tailed probability distribution
contribute to it.
We also find the property of temporal correlation for 2014 is different from that for 2015 and 2016.
This finding is verified by the differences in the variability of $h(q)$ and $f(\alpha)$, and
$\Delta h_{corr}$ and $R$.
The results of the autocorrelation of returns,
namely, the different autocorrelations at 1-min time lags between 2014 and other years, also supports this finding.
Ref.\cite{bariviera2017some} found that the Hurst exponent changes in 2014, 
which indicates properties of time series change. 
Although the difference between 2014 and other years could be related to this change of the Hurst exponent,  
we do not find any facts that cause this change.
By examining the market efficiency of Bitcoin with the Hurst exponent, 
Ref.\cite{urquhart2016inefficiency} finds that the Hurst exponent is less than 0.5, 
and concludes that Bitcoin time series is ant-persistence, and thus, inefficient.
On the other hand, by analyzing a power transformed Bitcoin returns, 
Ref.\cite{nadarajah2017inefficiency} claims that the transformed Bitcoin returns satisfy the efficient market hypothesis.
Our analysis on Bitcoin reveals that, for the full sample period (2014--2016), the Hurst exponent $h(2)$ is less than 0.5, and
Bitcoin time series shows the multifractal properties. 
Therefore, our findings support the conclusion of Ref.\cite{urquhart2016inefficiency}; that is, the inefficiency of Bitcoin.

The influence of "Brexit" on June 3, 2016 to the GBP--USD exchange rate and Bitcoin is analyzed in multifractality.
We find that Brexit influences the multifractality of the GBP--USD exchange rate, but no influence is seen 
in the multifractality of Bitcoin. Thus, Bitcoin was robust to Brexit. 

\section*{Acknowledgment}
Numerical calculations for this study were carried out at the
Yukawa Institute Computer Facility and facilities of the Institute of Statistical Mathematics.
The author thanks the anonymous referee for letting him know a recent  
study on the multifractality of the Bitcoin market submitted during the review process\cite{lahmiri2018chaos}.

\section*{References}

\bibliography{bitcoin}

\end{document}